\documentclass{vldb}

\makeatletter

\let\proof\@undefined
\let\endproof\@undefined
\makeatother

\usepackage{amsmath,amssymb,amstext} % Lots of math symbols and environments
\usepackage{graphicx} % For including graphics N.B. pdftex graphics driver 

\usepackage{adjustbox}
\usepackage{algorithm,algorithmicx, algpseudocode}
\usepackage{array}
\usepackage{amsfonts}
\usepackage[justification=centering]{caption}
\DeclareCaptionType{copyrightbox}
\usepackage{cite}
\usepackage{enumerate}
\usepackage{etoolbox}
\usepackage{float}
\usepackage{hyperref}
\usepackage{ifthen}
\usepackage{longtable}
\usepackage{mathtools}
\usepackage{mdframed}
\usepackage{multirow}
\usepackage{nicefrac}
\usepackage{paralist}
\usepackage{relsize}
\usepackage[justification=centering]{subcaption}
\usepackage{stmaryrd}
\usepackage{tikz}
\usepackage{tikz-qtree}
\usepackage{times}
\usepackage{xspace}

\usepackage{arydshln}

\hypersetup{
	pdfauthor={G\"{u}ne\c{s} Alu\c{c}},
	pdfcreator={},
}

\usetikzlibrary{arrows,calc,decorations.pathmorphing,decorations.pathreplacing,fit,patterns,positioning,shapes.geometric,shapes.multipart,snakes}
\newtheorem{theorem}{Theorem}

\newtheorem{corollary}[theorem]{Corollary}
\newtheorem{proof}{Proof}
\newtheorem{proofSketch}{Proof Sketch}
\newtheorem{definition}{Definition}

\newcommand{\record}[1]{\vec{#1}}

\newcolumntype{R}[2]{%
    >{\adjustbox{angle=#1,lap=\width-(#2)}\bgroup}%
    l%
    <{\egroup}%
}
\newcommand*\rot{\multicolumn{1}{R{30}{1em}}}% no optional argument here, please!

\newcommand{\intRange}[2]{
	\mathbb{Z}_{#1 \, \cdots \, #2}
}

\newcommand{\intRangeCart}[3]{
	\intRange{#1}{#2}^{#3}
}

\newcommand{\tunableLSH}{\textsc{Tu\-na\-ble-LSH}\xspace}

\newcommand{\symbolsTable}{
{\scriptsize
\setlength{\tabcolsep}{3pt}
\begin{tabular}{c r l}
	& \emph{Symbol}				& \emph{Description} \\ \hline
\multirow{5}{*}{\rotatebox[origin=c]{90}{Constants}}
	& $\omega$				& database size (i.e., number of records) \\
	& $\epsilon$				& number of pages in the storage system \\
	& $k$					& maximum no.~of query access vectors that can be stored \\
	& $b$					& number of entries in each record utilization counter \\
	& $t$					& current time \\ \hline
\multirow{10}{*}{\rotatebox[origin=c]{90}{Data structures}}
	& $\vec{q}$				& query access vector (contains $\omega$ bits) \\
	& $\vec{r}$				& record utilization vector (contains $k$ bits) \\
	& $\vec{c}$				& record utilization counter (contains $b$ entries) \\
	& $\vec{P}$				& depending on the context, a point in a $k$-dimensional \\
	&					& or $b$-dimensional (Taxicab) space \\
	& $M_{\omega \times k}$			& query access matrix; contains the last $k$ most \\
	&					& representative query access vectors (in columns), \\
	&					& or equivalently, $\omega$ record utilization vectors (in rows) \\
	& $C_{\omega \times b}$			& frequency matrix; represents record utilization frequency \\
	&					& over $b$ groups of query access vectors \\ \hline
\multirow{6}{*}{\rotatebox[origin=c]{90}{Accessors}}
	& $q[i]$				& value of the $i^{\text{th}}$ bit in query access vector $\vec{q}$ \\
	& $r[i]$				& value of the $i^{\text{th}}$ bit in record utilization vector $\vec{r}$ \\
	& $c[i]$				& value of the $i^{\text{th}}$ entry in record utilization counter $\vec{c}$ \\
	& $P[i]$				& value of the $i^{\text{th}}$ coordinate in point $\vec{P}$ \\
	& $M[i][j]$				& value of the $i^{\text{th}}$ row and $j^{\text{th}}$ column in matrix \\
	& $C[i][j]$				& value of the $i^{\text{th}}$ row and $j^{\text{th}}$ column in matrix \\ \hline
\multirow{4}{*}{\rotatebox[origin=c]{90}{Distances}}
	& $\delta(\vec{r_{x}}, \vec{r_{y}})$	& Hamming distance between two record utilization \\
	&					& vectors \\
	& $\delta^H(\vec{q_{x}}, \vec{q_{y}})$	& \textsc{min-hash} distance between two query access vectors\\
	& $\delta^M(\vec{P_{x}}, \vec{P_{y}})$	& Manhattan distance between two points \\ \hline
\end{tabular}
}
}

\newcommand{\dataStructuresTable}{
{\scriptsize
\begin{tabular}{r l}
\emph{Symbol}		& \emph{Description} \\ \hline
$\text{begin}$		& natural number between $0 \ldots (k-1)$, initial value is $0$ \\
$\text{size}$		& natural number between $0 \ldots (k-1)$, keeps track of \\
			& the number of query access vectors that are \\
			& currently being maintained, initial value is $0$ \\
$H_{k \times ?}$ 	& matrix that contains \textsc{min-hash} values \\
			& for each query access vector \\
$S[]$ 			& array of \emph{vector}(s), one for each MDS query point, that \\
			& pairs each MDS query point with a \emph{random} subset of points \\
$N[]$			& array of \emph{max-heap}(s), one for each MDS query point, that \\
			& pairs each MDS query point with a set of \emph{neighboring} points \\
$X[]$			& array of \emph{float}(s), represents the coordinate \\
			& (single dimensional) of each MDS query point \\
$V[]$			& array of \emph{float}(s), represents the current \\
			& (directional) velocity of each MDS query point \\ \hline
\end{tabular}
}
}

\def\VR{\kern-\arraycolsep\strut\vrule &\kern-\arraycolsep}
\def\vr{\kern-\arraycolsep & \kern-\arraycolsep}

\newcommand{\matrixOriginal}{
\let\quad\thinspace 
\scriptsize
\bordermatrix{
~ & q_{0} & q_{1} & q_{2} & q_{3} & q_{4} & q_{5} & q_{6} & q_{7} \cr
r_{0} & 0 & \textbf{1} & \textbf{1} & \textbf{1} & : \textbf{1} & \textbf{1} & \textbf{1} & \textbf{1} \cr
r_{1} & \textbf{1} & 0 & \textbf{1} & 0 & : 0 & 0 & 0 & 0 \cr \cdashline{2-9}
r_{2} & 0 & 0 & 0 & \textbf{1} & : 0 & \textbf{1} & 0 & \textbf{1} \cr
r_{3} & \textbf{1} & 0 & \textbf{1} & 0 & : \textbf{1} & 0 & \textbf{1} & 0 \cr \cdashline{2-9}
r_{4} & \textbf{1} & 0 & \textbf{1} & 0 & : \textbf{1} & 0 & \textbf{1} & 0 \cr
r_{5} & 0 & \textbf{1} & 0 & \textbf{1} & : 0 & \textbf{1} & 0 & \textbf{1} \cr \cdashline{2-9}
r_{6} & \textbf{1} & \textbf{1} & \textbf{1} & \textbf{1} & : \textbf{1} & \textbf{1} & \textbf{1} & \textbf{1} \cr
r_{7} & 0 & 0 & 0 & 0 & : 0 & 0 & 0 & 0 \cr
}
}

\newcommand{\matrixRowClustered}{
\let\quad\thinspace
\scriptsize
\bordermatrix{
~ & q_{0} & q_{1} & q_{2} & q_{3} & q_{4} & q_{5} & q_{6} & q_{7} \cr
r_{7} & 0 & 0 & 0 & 0 & : 0 & 0 & 0 & 0 \cr
r_{1} & \textbf{1} & 0 & \textbf{1} & 0 & : 0 & 0 & 0 & 0 \cr \cdashline{2-9}
r_{4} & \textbf{1} & 0 & \textbf{1} & 0 & : \textbf{1} & 0 & \textbf{1} & 0 \cr
r_{3} & \textbf{1} & 0 & \textbf{1} & 0 & : \textbf{1} & 0 & \textbf{1} & 0 \cr \cdashline{2-9}
r_{6} & \textbf{1} & \textbf{1} & \textbf{1} & \textbf{1} & : \textbf{1} & \textbf{1} & \textbf{1} & \textbf{1} \cr
r_{0} & 0 & \textbf{1} & \textbf{1} & \textbf{1} & : \textbf{1} & \textbf{1} & \textbf{1} & \textbf{1} \cr \cdashline{2-9}
r_{5} & 0 & \textbf{1} & 0 & \textbf{1} & : 0 & \textbf{1} & 0 & \textbf{1} \cr
r_{2} & 0 & 0 & 0 & \textbf{1} & : 0 & \textbf{1} & 0 & \textbf{1} \cr
}
}

\newcommand{\matrixColumnClustered}{
\let\quad\thinspace 
\scriptsize
\bordermatrix{
~ & q_{0} & q_{2} & q_{6} & q_{4} & q_{3} & q_{5} & q_{7} & q_{1} \cr
r_{7} & 0 & 0 & 0 & 0 & : 0 & 0 & 0 & 0 \cr
r_{1} & \textbf{1} & \textbf{1} & 0 & 0 & : 0 & 0 & 0 & 0 \cr \cdashline{2-9}
r_{4} & \textbf{1} & \textbf{1} & \textbf{1} & \textbf{1} & : 0 & 0 & 0 & 0 \cr
r_{3} & \textbf{1} & \textbf{1} & \textbf{1} & \textbf{1} & : 0 & 0 & 0 & 0 \cr \cdashline{2-9}
r_{6} & \textbf{1} & \textbf{1} & \textbf{1} & \textbf{1} & : \textbf{1} & \textbf{1} & \textbf{1} & \textbf{1} \cr
r_{0} & 0 & \textbf{1} & \textbf{1} & \textbf{1} & : \textbf{1} & \textbf{1} & \textbf{1} & \textbf{1} \cr \cdashline{2-9}
r_{5} & 0 & 0 & 0 & 0 & : \textbf{1} & \textbf{1} & \textbf{1} & \textbf{1} \cr
r_{2} & 0 & 0 & 0 & 0 & : \textbf{1} & \textbf{1} & \textbf{1} & 0 \cr
}
}

\newcommand{\matrixUpdated}{
\let\quad\thinspace
\scriptsize
\bordermatrix{
~ & q_{4} & q_{5} & q_{6} & q_{7} & q_{8} & q_{9} & q_{10} & q_{11} \cr
r_{7} & 0 & 0 & 0 & 0 & : 0 & 0 & 0 & 0 \cr \cdashline{2-9}
r_{1} & 0 & 0 & 0 & 0 & : 0 & 0 & 0 & 0\cr
r_{4} & \textbf{1} & 0 & \textbf{1} & 0 & : 0 & 0 & 0 & 0 \cr \cdashline{2-9}
r_{3} & \textbf{1} & 0 & \textbf{1} & 0 & : 0 & \textbf{1} & \textbf{1} & 0 \cr 
r_{6} & \textbf{1} & \textbf{1} & \textbf{1} & \textbf{1} & : \textbf{1} & \textbf{1} & \textbf{1} & \textbf{1} \cr \cdashline{2-9}
r_{0} & \textbf{1} & \textbf{1} & \textbf{1} & \textbf{1} & : \textbf{1} & \textbf{1} & \textbf{1} & \textbf{1} \cr 
r_{5} & 0 & \textbf{1} & 0 & \textbf{1} & : \textbf{1} & 0 & 0 & \textbf{1} \cr \cdashline{2-9}
r_{2} & 0 & \textbf{1} & 0 & \textbf{1} & : 0 & 0 & 0 & 0 \cr
}
}

\newcommand{\matrixOriginalTransformed}{
\let\quad\thinspace 
\scriptsize
\bordermatrix{
~ & G_{0} & G_{1} \cr
c_{7} & 0 & 0 \cr
c_{1} & 2 & 0 \cr \cdashline{2-3}
c_{4} & 2 & 2 \cr
c_{3} & 2 & 2 \cr \cdashline{2-3}
c_{6} & 4 & 4 \cr
c_{0} & 3 & 4 \cr  \cdashline{2-3}
c_{5} & 2 & 2 \cr
c_{2} & 1 & 2 \cr
}
}

\newcommand{\matrixCCTransformed}{
\let\quad\thinspace 
\scriptsize
\bordermatrix{
~ & G_{0} & G_{1} \cr
c_{7} & 0 & 0 \cr
c_{1} & 2 & 0 \cr \cdashline{2-3}
c_{4} & 4 & 0 \cr
c_{3} & 4 & 0 \cr \cdashline{2-3}
c_{6} & 4 & 4 \cr
c_{0} & 3 & 4 \cr \cdashline{2-3}
c_{5} & 0 & 4 \cr 
c_{2} & 0 & 3 \cr
}
}

\newcommand{\counterShifting}{
\setlength{\tabcolsep}{2pt}
\begin{tabular}{l | c c c c c c |}
					& \textbf{0}	& \textbf{1} 	& \textbf{2} 	& \textbf{3} 	& \textbf{4} 	& \textbf{5} 	\\ \hline
$t=0$					& $\Box$	& $\Box$	& $\Box$	& $\emptyset$	&		&		\\
$t=\lceil {}^{k} / {}_3 \rceil$		&		& $\Box$	& $\Box$	& $\Box$	& $\emptyset$	&		\\
$t=\lceil {}^{2k} / {}_3 \rceil$	&		&		& $\Box$	& $\Box$	& $\Box$	& $\emptyset$	\\
$t=k$					& $\emptyset$	& 		&		& $\Box$	& $\Box$	& $\Box$	\\
$t=\lceil {}^{4k} / {}_3 \rceil$		& $\Box$	& $\emptyset$	& 		&		& $\Box$	& $\Box$	\\
$t=\lceil {}^{5k} / {}_3 \rceil$		& $\Box$	& $\Box$	& $\emptyset$	& 		&		& $\Box$	\\
\end{tabular}
}

\title{Clustering RDF Databases Using Tunable-LSH}

\author{
{G\"une\c{s} Alu\c{c},  M. Tamer \"{O}zsu, and Khuzaima Daudjee}
\vspace{1.6mm}\\
\fontsize{10}{10}\selectfont\itshape
University of Waterloo David R. Cheriton School of Computer Science\\
\fontsize{9}{9}\selectfont\ttfamily\upshape
\{galuc,tamer.ozsu,kdaudjee\}@uwaterloo.ca\\
}
\begin{document}

\maketitle
\begin{abstract}
The Resource Description Framework (RDF) is a W3C standard for representing graph-structured data, and SPARQL is the standard query language for RDF. Recent advances in Information Extraction, Linked Data Management and the Semantic Web have led to a rapid increase in both the volume and the variety of RDF data that are publicly available. As businesses start to capitalize on RDF data, RDF data management systems are being exposed to workloads that are far more diverse and dynamic than what they were designed to handle. Consequently, there is a growing need for developing workload-adaptive and self-tuning RDF data management systems. To realize this vision, we introduce a fast and efficient method for dynamically clustering records in an RDF data management system. Specifically, we assume nothing about the workload upfront, but as SPARQL queries are executed, we keep track of records that are co-accessed by the queries in the workload and physically cluster them. To decide dynamically (hence, in constant-time) where a record needs to be placed in the storage system, we develop a new locality-sensitive hashing (LSH) scheme, \tunableLSH. Using \tunableLSH, records that are co-accessed across similar sets of queries can be hashed to the same or nearby physical pages in the storage system. What sets \tunableLSH apart from existing LSH schemes is that it can auto-tune to achieve the aforementioned clustering objective with high accuracy even when the workloads change. Experimental evaluation of \tunableLSH in our prototype RDF data management system, \emph{chameleon-db}, as well as in a standalone hashtable shows significant end-to-end improvements over existing solutions.
\end{abstract}

\section{Introduction}
\label{sec:introduction}
Physical data organization plays an important role in the performance tuning of database management systems.
A particularly important problem is clustering (in the storage system) records that are frequently co-accessed by queries in a workload.
Suboptimal clustering has negative performance implications due to random I/O and cache stalls~\cite{AilamakiVLDB1999}.
This problem has received attention in the context of SQL databases 
and has led to the introduction of tuning advisors that  work either in an \emph{offline}~\cite{AgrawalVLDB2000,ZilioVLDB2004} or \emph{online} fashion (i.e., self-tuning databases)~\cite{ChaudhuriVLDB2007}.

In this paper, we address the problem in the context of RDF data management systems.
SPARQL workloads are far more dynamic than SQL workloads~\cite{AriasUSEWOD2011,KirchbergUSEWOD2011};
	yet, tuning techniques for RDF data management systems are in their infancy,
	and relational solutions are not directly applicable.
More specifically, depending on the workload,
	it might be necessary to completely change
	the underlying physical representation in an RDF data management system,
	such as by dynamically switching from a row-oriented representation to a columnar representation~\cite{AlucPVLDB2014}.
On the other hand, existing online tuning techniques work well only when the schema changes are minor~\cite{BrunoVLDB2006}.
Consequently, with the increasing demand to support highly dynamic workloads in RDF~\cite{AriasUSEWOD2011,KirchbergUSEWOD2011},
	there is a growing need to develop more adaptive tuning solutions,
	in which records in an RDF database can be dynamically and continuously clustered
	based on the current workload.

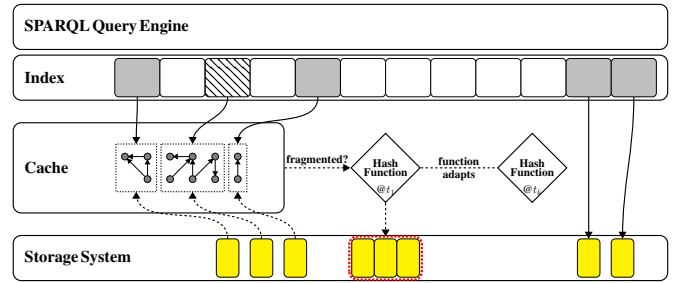
\begin{figure}[t]
	\centering
     \scalebox{0.3}{\begin{tikzpicture}
	\newcommand{\boxWidth}{29cm}
	\newcommand{\boxHeight}{2cm}
	\newcommand{\smallBoxWidth}{12cm}
	\newcommand{\textMargin}{1cm}

	\node (b1) [draw, 
			rounded corners=10pt,
			line width=0.5mm,
			minimum width=\boxWidth, 
			minimum height=\boxHeight, 
			text width={\boxWidth-\textMargin}, 
			align=left] 
				{\huge \textbf{Storage System}};

	\node (b2) [draw, 
			above of=b1, 
			xshift={(\smallBoxWidth-\boxWidth)/2}, 
			yshift={1.75cm + \boxHeight/2+0.25cm}, 
			rounded corners=10pt,
			line width=0.5mm,
			minimum width=\smallBoxWidth, 
			minimum height={2*\boxHeight}, 
			text width={\smallBoxWidth-\textMargin}, 
			align=left] 
				{\huge \textbf{Cache}};

	\node (b3) [draw,
			above of=b2,
			xshift={(\boxWidth-\smallBoxWidth)/2},
			yshift={1.75cm + \boxHeight/2+0.25cm},
			rounded corners=10pt,
			line width=0.5mm,
			minimum width=\boxWidth,
			minimum height=\boxHeight,
			text width={\boxWidth-\textMargin},
			align=left]
				{\huge \textbf{Index}};

	\node (b4) [draw,
			above of=b3,
			yshift={\boxHeight/2+0.25cm},
			rounded corners=10pt,
			line width=0.5mm,
			minimum width=\boxWidth,
			minimum height=\boxHeight,
			text width={\boxWidth-\textMargin},
			align=left]
				{\huge \textbf{SPARQL Query Engine}};

	%%% Draw the records inside the storage system box...
	\node (r1) [draw, left of=b1, xshift=-4cm, rounded corners=5pt, minimum width=1cm, minimum height={\boxHeight-0.3cm}, fill=yellow] {};
	\node (r2) [draw, left of=b1, xshift=-2.5cm, rounded corners=5pt, minimum width=1cm, minimum height={\boxHeight-0.3cm}, fill=yellow] {};
	\node (r3) [draw, left of=b1, xshift=-1cm, rounded corners=5pt, minimum width=1cm, minimum height={\boxHeight-0.3cm}, fill=yellow] {};
	\node (r4) [draw, left of=b1, xshift=2cm, rounded corners=5pt, minimum width=1cm, minimum height={\boxHeight-0.3cm}, fill=yellow] {};
	\node (r5) [draw, left of=b1, xshift=3cm, rounded corners=5pt, minimum width=1cm, minimum height={\boxHeight-0.3cm}, fill=yellow] {};
	\node (r6) [draw, left of=b1, xshift=4cm, rounded corners=5pt, minimum width=1cm, minimum height={\boxHeight-0.3cm}, fill=yellow] {};
	\node (r7) [draw, left of=b1, xshift=12cm, rounded corners=5pt, minimum width=1cm, minimum height={\boxHeight-0.3cm}, fill=yellow] {};
	\node (r8) [draw, left of=b1, xshift=13.5cm, rounded corners=5pt, minimum width=1cm, minimum height={\boxHeight-0.3cm}, fill=yellow] {};

	%%% Draw the dashed rectangle around r4, r5 and r6...
	\draw [red, rounded corners=10pt, line width=1mm, dotted] ($(r4.south west) + (-0.1cm, -0.1cm)$) rectangle ($(r6.north east) + (0.1cm, 0.1cm)$);

	%%% Draw the adapting hash function...
	\node (hf1) [draw, 
			diamond, 
			right of=b2,
			line width=0.5mm,
			xshift={(\smallBoxWidth/2)+3.5cm}, 
			align=center] 
				{\Large \textbf{Hash} \\ \Large \textbf{Function}};
	\node [below of=hf1] {\Large @$t_{1}$};

	\node (hf2) [draw,
			diamond,
			right of=hf1,
			line width=0.5mm,
			xshift=5.5cm,
			align=center]
				{\Large \textbf{Hash} \\ \Large \textbf{Function}};
	\node [below of=hf2] {\Large @$t_{k}$};

	%%% Draw the records inside the index box...
	\node (p1) [draw, left of=b3, xshift=-8cm, rounded corners=5pt, minimum width=2cm, minimum height={\boxHeight-0.3cm}, fill=lightgray] {};
	\node (p2) [draw, left of=b3, xshift=-6cm, rounded corners=5pt, minimum width=2cm, minimum height={\boxHeight-0.3cm}, fill=white] {};
	\node (p3) [draw, left of=b3, xshift=-4cm, rounded corners=5pt, minimum width=2cm, minimum height={\boxHeight-0.3cm}, pattern=north west lines, pattern color=black] {};
	\node (p4) [draw, left of=b3, xshift=-2cm, rounded corners=5pt, minimum width=2cm, minimum height={\boxHeight-0.3cm}, fill=white] {};
	\node (p5) [draw, left of=b3, xshift=0cm, rounded corners=5pt, minimum width=2cm, minimum height={\boxHeight-0.3cm}, fill=lightgray] {};
	\node (p6) [draw, left of=b3, xshift=2cm, rounded corners=5pt, minimum width=2cm, minimum height={\boxHeight-0.3cm}, fill=white] {};
	\node (p7) [draw, left of=b3, xshift=4cm, rounded corners=5pt, minimum width=2cm, minimum height={\boxHeight-0.3cm}, fill=white] {};
	\node (p8) [draw, left of=b3, xshift=6cm, rounded corners=5pt, minimum width=2cm, minimum height={\boxHeight-0.3cm}, fill=white] {};
	\node (p9) [draw, left of=b3, xshift=8cm, rounded corners=5pt, minimum width=2cm, minimum height={\boxHeight-0.3cm}, fill=white] {};
	\node (p10) [draw, left of=b3, xshift=10cm, rounded corners=5pt, minimum width=2cm, minimum height={\boxHeight-0.3cm}, fill=white] {};
	\node (p11) [draw, left of=b3, xshift=12cm, rounded corners=5pt, minimum width=2cm, minimum height={\boxHeight-0.3cm}, fill=lightgray] {};
	\node (p12) [draw, left of=b3, xshift=14cm, rounded corners=5pt, minimum width=2cm, minimum height={\boxHeight-0.3cm}, fill=lightgray] {};

	%%% Draw the graphs in cache...
	
	\node (v11) [draw, circle, left of=b2, xshift=-0.05cm, yshift=0.5cm, fill=gray, minimum size=0.3cm, inner sep=0cm] {};
	\node (v12) [draw, circle, right of=v11, fill=gray, minimum size=0.3cm, inner sep=0cm] {};
	\node (v13) [draw, circle, below of=v12, fill=gray, minimum size=0.3cm, inner sep=0cm] {};
	\path [draw=black, solid, line width=0.1mm, fill=black, -triangle 60] (v12) -- (v11);
	\path [draw=black, solid, line width=0.1mm, fill=black, -triangle 60] (v13) -- (v11);
	\path [draw=black, solid, line width=0.1mm, fill=black, -triangle 60] (v13) -- (v12);
	\draw [line width=0.5mm, dotted] ($(v11.north) + (-0.4cm, 0.4cm)$) rectangle ($(v13.south) + (0.4cm, -0.4cm)$);

	\node (v21) [draw, circle, right of=v12, fill=gray, minimum size=0.3cm, inner sep=0cm] {};
	\node (v22) [draw, circle, below of=v21, fill=gray, minimum size=0.3cm, inner sep=0cm] {};
	\node (v23) [draw, circle, right of=v21, fill=gray, minimum size=0.3cm, inner sep=0cm] {};
	\node (v24) [draw, circle, below of=v23, fill=gray, minimum size=0.3cm, inner sep=0cm] {};
	\node (v25) [draw, circle, right of=v23, fill=gray, minimum size=0.3cm, inner sep=0cm] {};
	\node (v26) [draw, circle, below of=v25, fill=gray, minimum size=0.3cm, inner sep=0cm] {};
	\path [draw=black, solid, line width=0.1mm, fill=black, -triangle 60] (v23) -- (v21);
	\path [draw=black, solid, line width=0.1mm, fill=black, -triangle 60] (v22) -- (v23);
	\path [draw=black, solid, line width=0.1mm, fill=black, -triangle 60] (v24) -- (v23);
	\path [draw=black, solid, line width=0.1mm, fill=black, -triangle 60] (v24) -- (v25);
	\path [draw=black, solid, line width=0.1mm, fill=black, -triangle 60] (v25) -- (v26);
	\draw [line width=0.5mm, dotted] ($(v21.north) + (-0.4cm, 0.4cm)$) rectangle ($(v26.south) + (0.4cm, -0.4cm)$);

	\node (v31) [draw, circle, right of=v25, fill=gray, minimum size=0.3cm, inner sep=0cm] {};
	\node (v32) [draw, circle, below of=v31, fill=gray, minimum size=0.3cm, inner sep=0cm] {};
	\path [draw=black, solid, line width=0.1mm, fill=black, -triangle 60] (v32) -- (v31);
	\draw [line width=0.5mm, dotted] ($(v31.north) + (-0.4cm, 0.4cm)$) rectangle ($(v32.south) + (0.4cm, -0.4cm)$);

	%%% Place arrows and pointers...
	\path [draw=black, solid, dashed, line width=0.5mm, -triangle 60] (r1.north) to[out=90, in=270, looseness=1] ($(v13.south) + (-0.5cm, -0.4cm)$);
	\path [draw=black, solid, dashed, line width=0.5mm, -triangle 60] (r2.north) to[out=90, in=270, looseness=1] ($(v24.south) + (0cm, -0.4cm)$);
	\path [draw=black, solid, dashed, line width=0.5mm, -triangle 60] (r3.north) to[out=90, in=270, looseness=1] ($(v32.south) + (0cm, -0.4cm)$);

	\path [draw=black, solid, dashed, line width=0.5mm, -triangle 60] (b2.east) -- (hf1.west) node[midway, above] {\Large \textbf{fragmented?}};
	\path [draw=black, solid, dashed, line width=0.5mm, -triangle 60] (hf1.south) to[out=270, in=90, looseness=1] ($(r5.north) + (0cm, 0.1cm)$);

	\path [draw=black, solid, dashed, line width=0.5mm] (hf1.east) -- (hf2.west) node[midway, above] {\Large \textbf{function}} node[midway, below] {\Large \textbf{adapts}};

	\path [draw=black, solid, line width=0.5mm, -triangle 60] (p1.south) to[out=270, in=90, looseness=1] ($(v11.north) + (0.5cm, 0.4cm)$); 
	\path [draw=black, solid, line width=0.5mm, -triangle 60] (p3.south) to[out=270, in=90, looseness=1] ($(v23.north) + (0cm, 0.4cm)$); 
	\path [draw=black, solid, line width=0.5mm, -triangle 60] (p5.south) to[out=270, in=90, looseness=1] ($(v31.north) + (0cm, 0.4cm)$); 
	
	\path [draw=black, solid, line width=0.5mm, -triangle 60] (p11.south) to[out=270, in=90, looseness=1] (r7.north);
	\path [draw=black, solid, line width=0.5mm, -triangle 60] (p12.south) to[out=270, in=90, looseness=1] (r8.north);

\end{tikzpicture}}  

	\caption{Adaptive record placement using a combination of adaptive hashing and caching~\cite{AlucPVLDB2014}.}
	\label{fig:adaptive-record-placement}
\end{figure}

Whenever a SPARQL query is executed, 
	there is an opportunity to observe how records in an RDF database are being utilized.
This information about query access patterns can be used to dynamically cluster records in the storage system.
Dynamism is important in RDF systems because of the high variability and dynamism in SPARQL workloads~\cite{AriasUSEWOD2011,KirchbergUSEWOD2011}.
While this problem has been studied as physical clustering~\cite{Lightstone2007} and distribution design~\cite{CeriIEEETSE1983},
	the highly dynamic nature of the queries over RDF data introduces new challenges.
First, traditional algorithms are offline, and 
	since clustering is an NP-hard problem and most approximations have quadratic complexity~\cite{JainCOMPSUR1999},
	they are not suitable for online database clustering.
Instead, techniques are needed with similar clustering objectives, but that have constant running time.
Second, systems are typically expected to execute most queries in subseconds~\cite{NahBaIT2004}, 
	leaving only fractions of a second to update their physical data structures
	(i.e., in our case, we are concerned with dynamically moving records across the storage system).

We address the aforementioned issues by making two contributions.
First, as shown in Fig.~\ref{fig:adaptive-record-placement}, instead of clustering the whole database,
	we cluster only the ``warm" portions of the database by relying on the admission policy of the existing database cache.
Second, we develop a self-tuning locality-sensitive hash (LSH) function, namely, \tunableLSH
	to decide in constant-time where in the storage system to place a record.
\tunableLSH has two important properties:
	\begin{itemize}
		\item	It tries to ensure that
				\begin{inparaenum}[(i)]
					\item records with \emph{similar} utilization patterns (i.e., those records 
						that are co-accessed across similar sets of queries) 
						are mapped as much as possible to the same pages in the storage system, while
					\item minimizing the number of records with \emph{dissimilar} utilization patterns that are falsely mapped to the same page.
				\end{inparaenum}
		\item Unlike conventional LSH~\cite{IndykSTOC98,GionisVLDB1999}, \tunableLSH can auto-tune so as to
				achieve the aforementioned clustering objectives with high accuracy
				even when the workloads change.
	\end{itemize}

These ideas are illustrated in Fig.~\ref{fig:adaptive-record-placement}.
Let us assume that initially, the records in a database are \emph{not} clustered according to any particular workload.
Therefore, the performance of the system is suboptimal.
However, every time records are fetched from the storage system, there is an opportunity to
	bring together into a single page those records that are co-accessed but are fragmented across the storage system.
\tunableLSH achieves these with minimal overhead.
Furthermore, \tunableLSH is continuously updated to reflect any changes in the workload characteristics.
Consequently, as more queries are executed, records in the database become more clustered, and 
	the performance of the system improves.

The paper is organized as follows:
Section~\ref{sec:related-work} discusses related work.
Section~\ref{sec:problem} gives a conceptual description of the problem.
Section~\ref{sec:naive} describes the overview of our approach while Section~\ref{sec:details} provides the details.
In Section~\ref{sec:evaluation}, we describe how physical clustering takes place in the database, in particular,	
	how \tunableLSH can be used in an RDF data management system, and we experimentally evaluate our techniques.
Finally, we discuss conclusions and future work in Section~\ref{sec:conclusions}.

\section{Related Work}
\label{sec:related-work}
Locality-sensitive hashing (LSH)~\cite{IndykSTOC98,GionisVLDB1999} has been used in various contexts such as nearest neighbour search~\cite{IndykSTOC98,FerhatosmanogluICDE2001,HouleICDE2005,AndoniFOCS2006,AthitsosICDE2008,TaoTODS2010}, Web document clustering~\cite{BroderSEQUENCES1997,BroderJCSS2000} and query plan caching~\cite{AlucICDE2012}. In this paper, we use LSH in the physical design of RDF databases. While multiple families of LSH functions have been developed~\cite{IndykSTOC98,GionisVLDB1999,BroderJCSS2000,CharikarSTOC2002,DatarSOCG2004}, these functions assume that the input distribution is either uniform or static. In contrast, \tunableLSH can continuously adapt to changes in the input distribution to achieve higher accuracy, which translates to adapting to changes in the query access patterns in the workloads in the context of RDF databases.

Physical design has been the topic of an ongoing discussion in the world of RDF and SPARQL~\cite{WilkinsonHPL2006,SidirourgosPVLDB2008,AbadiVLDBJ2009,AlucPVLDB2014}. One option is to represent data in a single large table~\cite{CarrollWWW2004} and build clustered indexes, where each index implements a different sort order~\cite{HarthASWC2007,WeissPVLDB2008,NeumannPVLDB2010}. It has also been argued that grouping data can help improve performance~\cite{SidirourgosPVLDB2008,AbadiVLDBJ2009}. For this reason, multiple physical representations have been developed: 
in the \emph{group-by-predicate} representation, the database is vertically partitioned and the tables are stored in a column-store~\cite{AbadiVLDBJ2009};
in the \emph{group-by-entity} representation, implicit relationships within the database are discovered (either manually~\cite{WilkinsonHPL2006} or automatically~\cite{BorneaSIGMOD2013}), and the RDF data are mapped to a relational database; and in the \emph{group-by-vertex} representation, RDF's inherent graph-structure is preserved, whereby data can be grouped on the vertices in the graph~\cite{ZouPVLDB2011}. These workload-oblivious representations have issues for different types of queries, due to reasons such as fragmented data, unnecessarily large intermediate result tuples generated during query evaluation and/or suboptimal pruning by the indexes~\cite{AlucPVLDB2014}.

To address some of these issues, workload-aware techniques have been proposed~\cite{GoasdouePVLDB2011,HoseICDEWorkshops2013}. For example, view materialization techniques have been implemented for RDF over relational engines~\cite{GoasdouePVLDB2011}. However, these materialized views are difficult to adapt to changing workloads for reasons discussed in Section~\ref{sec:introduction}. Workload-aware distribution techniques have also been developed for RDF~\cite{HoseICDEWorkshops2013} and implemented in systems such as WARP~\cite{HoseICDEWorkshops2013} and Partout~\cite{GalarragaWWWCompanion2014}, but, these systems are not runtime-adaptive. With \tunableLSH, we aim to address the problem adaptively, by clustering fragmented records in the database based on the workload.

While there are self-tuning SQL databases~\cite{ChaudhuriVLDB2007,IdreosCIDR2007,IdreosPVLDB2011} and techniques for automatic schema design in SQL~\cite{BelloVLDB1998,AgrawalVLDB2000,ZilioVLDB2004,Lightstone2007}, these techniques are not directly applicable to RDF. In RDF, the advised changes to the underlying physical schema can be drastic, for example, requiring the system to switch from a row-oriented representation to a columnar one, all at runtime, which are hard to achieve using existing techniques. Consequently, there have been efforts in designing workload-adaptive and self-tuning RDF data management systems~\cite{PapailiouWWWCompanion2012,AlucUW2013,AlHarbiCoRR2014,AlucPVLDB2014,PapailiouSIGMOD2015}. In H2RDF~\cite{PapailiouWWWCompanion2012}, the choice between centralized versus distributed execution is made adaptively. In PHDStore~\cite{AlHarbiCoRR2014}, data are adaptively replicated and distributed across the compute nodes; however, the underlying physical layout is fixed within each node. A mechanism for adaptively caching partial results is introduced in~\cite{PapailiouSIGMOD2015}. With \tunableLSH, we are trying to address the adaptive record layout problem, therefore, we believe that \tunableLSH will complement existing techniques and facilitate the development of runtime adaptive RDF systems.

\section{Preliminaries}
\label{sec:problem}
Given a sequence of database records 
	that represent the records' serialization order in the storage system, 
	the access patterns of a query can conceptually be represented as a bit vector,
	where a bit is set to $1$ if the corresponding record in the sequence is accessed by the query.
We call this bit vector a \emph{query access vector} ($\vec{q}$).

Depending on the system, a record may denote a single RDF triple (i.e., the atomic unit of information in RDF), 
	as in systems like RDF-3x~\cite{NeumannVLDBJ2010},
	or a collection of RDF triples such as in chamele\-on-db~\cite{AlucUW2013,AlucICDE2015}.
Our conceptual model is applicable either way.

\begin{figure}[t]
	\centering
	\begin{subfigure}[t]{0.475\linewidth}
		\begin{center}
			\matrixOriginal
		\end{center}
		\caption{Representation at $t=8$.}
		\label{fig:matrix-rep:original}
	\end{subfigure}
	\hfill
	\begin{subfigure}[t]{0.475\linewidth}
		\begin{center}
			\matrixRowClustered
		\end{center}
		\caption{Clustered on rows}
		\label{fig:matrix-rep:rows}
	\end{subfigure}
	\begin{subfigure}[t]{0.475\linewidth}
		\begin{center}
			\matrixColumnClustered
		\end{center}
		\caption{Clustered on rows and columns}
		\label{fig:matrix-rep:columns}
	\end{subfigure}
	\hfill
	\begin{subfigure}[t]{0.2375\linewidth}
		\begin{center}
			\matrixOriginalTransformed
		\end{center}
		\caption{Grouping of bits}
		\label{fig:rows:transformed}
	\end{subfigure}
	\hfill
	\begin{subfigure}[t]{0.2375\linewidth}
		\begin{center}
			\matrixCCTransformed
		\end{center}
		\caption{Alternative grouping}
		\label{fig:rc:transformed}
	\end{subfigure}
	\caption{Matrix representation of query access patterns.}
	\label{fig:matrix-rep}
\end{figure}

As more queries are executed,
	their query access vectors can be accumulated column-by-column in a matrix,
	as shown in Fig.~\ref{fig:matrix-rep:original}.
We call this matrix a \emph{query access matrix}.
For presentation, let us assume that queries are numbered 
	according to their order of execution by the RDF data management system.

Each row of the query access matrix constitutes what we call a \emph{record utilization vector} ($\vec{r}$), 
	which represents the set of queries that access record $r$.
As a convention, to distinguish between a query and its access vector  (likewise, a record and its utilization vector),
	we  use the symbols $q$ and $\vec{q}$ (likewise, $r$ and $\vec{r}$), respectively.
The complete list of symbols are given in Table~\ref{tab:symbols}.

\begin{table}[t]
	\begin{center}
		\symbolsTable
	\end{center}
	\caption{Symbols used throughout the manuscript}
	\label{tab:symbols}
\end{table}

To model the memory hierarchy, 
	we use an additional notation in the matrix representation:
	records that are physically stored together
	on the same disk/memory page
	should be grouped together in the query access matrix.
For example, Fig.~\ref{fig:matrix-rep:original} and Fig.~\ref{fig:matrix-rep:rows} represent two alternative ways 
	in which the records in an RDF database can be clustered (groups are separated by horizontal dashed lines).
Even though both figures depict essentially the same query access patterns, 
	the physical organization in Fig.~\ref{fig:matrix-rep:rows} is preferable,
	because in Fig.~\ref{fig:matrix-rep:original}, most queries require access to $4$ pages each, whereas
	in Fig.~\ref{fig:matrix-rep:rows}, the number of accesses is reduced by almost half.

Given a sequence of queries and the number of pages in the storage system,
	our objective is to \emph{store records with similar utilization vectors together
	so as to minimize the total number of page accesses}.
To determine the similarity between record utilization vectors, 
	we rely on the following property.
Two records are co-accessed by a query if both of the corresponding bits in that query's access vector are set to $1$.
Extending this concept to a set of queries, we say that two records are co-accessed across multiple queries if the corresponding
	bits in the record utilization vectors are set to $1$ for all the queries in the set.
For example, according to Fig.~\ref{fig:matrix-rep:original},
	records $r_{1}$ and $r_{3}$ are co-accessed by queries $q_{0}$ and $q_{2}$, and
	records $r_{0}$ and $r_{6}$ are co-accessed across the queries $q_{1}$--$q_{7}$.

Given a sequence of queries, it may often be the case that a pair of records are not co-accessed in \emph{all} of the queries.
Therefore, to measure the extent to which a pair of records are co-accessed, we rely on their Hamming distance~\cite{Hamming1986}.
Specifically, given two record utilization vectors for the same sequence of queries, 
	their Hamming distance---denoted as $\delta(\vec{q}_{x}, \vec{q}_{y})$---is defined 
	as the minimum number of substitutions necessary to make the two bit vectors the same~\cite{Hamming1986}.\footnote{The Hamming
		distance between two record utilization vectors is equal to their edit distance~\cite{Levenshtein1966}, 
		as well as the Manhattan distance~\cite{Krause1986} between these two vectors in $l_{1}$ norm.}
Hence, the smaller the Hamming distance between a pair of records, the greater the extent to which they are co-accessed.

Consider the record utilization vectors $\record{r_{0}}$, $\record{r_{2}}$, $\record{r_{5}}$ and $\record{r_{6}}$ 
	across the query sequence $q_{0}$--$q_{7}$ in Fig.~\ref{fig:matrix-rep:original}.
The pairwise Hamming distances are as follows:
	$\delta( {r_{0}}, {r_{6}} ) = 1$,
	$\delta( {r_{2}}, {r_{5}} ) = 1$,
	$\delta( {r_{0}}, {r_{5}} ) = 3$,
	$\delta( {r_{0}}, {r_{2}} ) = 4$,
	$\delta( {r_{5}}, {r_{6}} ) = 4$ and
	$\delta( {r_{2}}, {r_{6}} ) = 5$.
Consequently, to achieve better physical clustering, we should try to store
	$r_{0}$ and $r_{6}$ together and $r_{2}$ and $r_{5}$ together, while keeping
	$r_{0}$ and $r_{6}$ apart from $r_{2}$ and $r_{5}$.

\section{Overview of Tunable-LSH}
\label{sec:naive}
Although we are dealing with a clustering problem, 
	the dynamic nature of queries over RDF data
	necessitate a solution different than existing ones~\cite{AlucPVLDB2014}.
That is, while conventional clustering algorithms~\cite{JainCOMPSUR1999} 
	might be perfectly applicable for the \emph{offline} tuning of a database,
	in an \emph{online} scenario, 
	even the most efficient algorithms may be impractical unless records
	are clustered on-the-fly and within microseconds.	
Clustering is an NP-complete problem~\cite{JainCOMPSUR1999}, and
	most approximations take at least quadratic time.
	It is not very well-understood which clustering algorithm is more suitable for which types of input distributions~\cite{AckermanNIPS2010}, 
	let alone the fact that incremental versions of these algorithms are largely domain-specific~\cite{Aggarwal2013}.
Therefore, we develop \tunableLSH, which is a self-tuning locality-sensitive hash (LSH) function.
As records are fetched from the storage system, we keep track of records that are fragmented.
Then, we use use \tunableLSH to decide, in constant-time,
	how a fragmented record needs to be clustered in the storage system (cf., Fig.~\ref{fig:adaptive-record-placement}).
Furthermore, we develop methods to continuously auto-tune this LSH function to
	adapt to changing query access patterns that we encounter while executing a workload.
This way, \tunableLSH can achieve much higher clustering accuracy than conventional LSH schemes, which are static.
%%%In the remainder of this section, we introduce \tunableLSH, and the algorithm that auto-tunes it.

Let $\intRange{\alpha}{\beta}$ denote the set of integers in the interval $[\alpha, \, \beta]$, and
	let $\intRangeCart{\alpha}{\beta}{n}$ denote the $n$-fold Cartesian product:
	\begin{align*}
		\underbrace{\intRange{\alpha}{\beta} \times \cdots \times \intRange{\alpha}{\beta}}_{n}.
	\end{align*}
Let us assume that we are given a non-injective, surjective function 
	$f: \intRange{0}{(k-1)} \rightarrow \intRange{0}{(b-1)}$, where
	$b \ll k$, and for all $y \in \intRange{0}{(b-1)}$, it holds that
	\begin{align*}
		\Big| \{ x : f(x) = y \} \Big| \leq \Big\lceil \frac{k}{b} \Big\rceil.
	\end{align*}
In other words, $f$ is a hash function with the property that,
	given $k$ input values and $b$ possible outcomes,
	no more than $\lceil \frac{k}{b} \rceil$ values in the domain of the function
	will be hashed to the same value.
Then, we define \tunableLSH as $h: \intRangeCart{0}{1}{k} \rightarrow \intRange{0}{(\epsilon-1)}$,
	where $\epsilon$ represents the number of pages in the storage system.
More specifically, $h$ is defined
	as a composition of two functions $h_{1}$ and $h_{2}$.

\begin{definition}[\tunableLSH] \label{def:tunable-lsh}
\begin{align*}
	\intertext{Let} 
	\vec{r} 	&= (r[0], \ldots, r[k\!-\!1]) \, \in \, \intRangeCart{0}{1}{k}, \: \text{and} \\
	\vec{c} 	&= (c[0], \ldots, c[b\!-\!1]) \, \in \, \intRangeCart{0}{\lceil \frac{k}{b} \rceil}{b}.
	\intertext{Then, a tunable LSH function $h$ is defined as}
	h &= \mathbf{h_{2}} \circ \mathbf{h_{1}} \\
	\intertext{where}
	\intertext{ 
			$\mathbf{h_{1}}: \intRangeCart{0}{1}{k} \rightarrow \intRangeCart{0}{\lceil \frac{k}{b} \rceil}{b}$, 
			where $h_{1}(\vec{r}) = \vec{c}$ iff 
	}
	\forall y \; c[y] &= \sum \limits_{x=0}^{k-1}
		\left\{
			\begin{array}{lr}
				r[x] & : f(x) = y \\
				0 & : f(x) \neq y
			\end{array}
		\right. \\
	\intertext{
			$\mathbf{h_{2}}: \intRangeCart{0}{\lceil \frac{k}{b} \rceil}{b} \rightarrow \intRange{0}{(\epsilon-1)}$,
			where $h_{2}(\vec{c}) = v$ iff
	}
	v &=
		\left\{
			\begin{array}{l}
				\text{coordinate of $\vec{c}$ (rounded to the} \\
				\text{nearest integer) on a space-filling} \\
				\text{curve~\cite{MortonIBM1966} of length $\epsilon$ that covers} \\
				\text{$\intRangeCart{0}{\lceil \frac{k}{b} \rceil}{b}$}
			\end{array}
		\right.
\end{align*}
\end{definition}

According to Def.~\ref{def:tunable-lsh}, $h$ is constructed as follows:
\begin{enumerate}
	\item Using a hash function $f$ (which can be treated as a black box for the moment),
			a record utilization vector $\vec{r}$ with $k$ bits
			is divided into $b$ disjoint segments $\vec{r}_{0}, \ldots, \vec{r}_{b-1}$ such that
			$\vec{r}_{0}, \ldots, \vec{r}_{b-1}$ contain all the bits in $\vec{r}$, and
			each $\vec{r}_{i} \in \{ \vec{r}_{0}, \ldots, \vec{r}_{b-1}\}$ has at most $\lceil \frac{k}{b} \rceil$ bits.
			Then, a record utilization counter $\vec{c}$ with $b$ entries is computed such that
			the $i^{\text{th}}$ entry of $\vec{c}$ (i.e., $c[i]$) 
			contains the number of $1$-bits in $\vec{r}_{i}$.
			Without loss of generality, a record utilization counter $\vec{c}$ can be represented 
			as	a $b$-dimensional point in the coordinate system $\intRangeCart{0}{\lceil \frac{k}{b} \rceil}{b}$.
	\item The final hash value is computed by ordering the points in $\intRangeCart{0}{\lceil \frac{k}{b} \rceil}{b}$
			using a space-filling curve~\cite{MortonIBM1966}.
\end{enumerate}

In Section~\ref{sec:details:properties}, we show that
	\tunableLSH that maps
	$k$-dimen\-sional record utilization vectors 
	to natural numbers in the interval $[0, \ldots, \epsilon-1]$ 
	is locality-sensitive,
	with two important implications:
	\begin{inparaenum}[(i)]
		\item records with similar record utilization vectors (i.e., small Hamming distances) are likely
				going to be hashed to the same value, while
		\item records with dissimilar record utilization vectors are likely
				going to be separated.
	\end{inparaenum}
Therefore, the problem of clustering records in the storage system
	can be approximated using \tunableLSH, 
	such that clustering $n$ records takes $\textsc{O}(n)$ time.

The quality of \tunableLSH, that is, 
	how well it approximates the original Hamming distances,
	depends on two factors:
	\begin{inparaenum}[(i)]
		\item the characteristics of the workload so far, which is reflected by the bit distribution in the record utilization vectors, and
		\item the choice of $f$.
	\end{inparaenum}
In Section~\ref{sec:details:adaptive}, we demonstrate that $f$ can be tuned
	to adapt to the changing patterns in record utilization vectors
	to maintain the approximation quality of \tunableLSH at a steady and high level.

\begin{algorithm}[t]
	\caption{Initialize} \label{alg:overview:initialize}
	{ \scriptsize
	\begin{algorithmic}[1]
		\Ensure
			\Statex Record utilization counters are allocated and initialized
		\Procedure{Initialize}{\null}
			\State construct int $C$[$\omega$][$2b$] 
				\Comment{\parbox[t]{.375\linewidth}{For simplicity, $C$ is allocated statically; however, in practice, it can be allocated dynamically to reduce memory footprint.}}
			\ForAll {$i$ $\in$ $(0, \ldots, \omega-1)$}
				\ForAll {$j$ $\in$ $(0, \ldots, 2b-1)$}
					\State $C$[$i$][$j$]	$\gets$ $0$
				\EndFor
			\EndFor
		\EndProcedure
	\end{algorithmic}
	}
\end{algorithm}

\begin{algorithm}[t]
	\caption{Tune} \label{alg:overview:tune}
	{ \scriptsize
	\begin{algorithmic}[1]
		\Require 
			\Statex $\vec{\mathbf{q_{t}}}$: query access vector produced at time $t$
		\Ensure
			\Statex Underlying data structures are updated and $f$ is tuned
						such that the LSH function maintains a steady approximation quality
		\Procedure{Tune}{$\vec{q_{t}}$}
			\State \Call{Reconfigure-F}{$\vec{q_{t}}$} \label{alg:overview:tune:la}
			\ForAll {$i$ $\in$ \Call{Positional}{$\vec{q_{t}}$}} \label{alg:overview:tune:lb}
				\State loc $\gets$ $f(t)$
				\If { loc $<$ $($shift $\% \: b)$ }
					\State loc \verb!+=! $b$
				\EndIf
				\State $C$[$i$][loc]\verb!++! \label{alg:overview:tune:l1}
					\Comment{\parbox[t]{.375\linewidth}}{Increment record utilization counters based on the new query access pattern}
				\If {$t \% \lceil \frac{k}{b} \rceil = 0$} \Comment{\parbox[t]{.375\linewidth}}{Reset ``old" counters}
					\State shift\verb!++!
					\State $C$[$i$][$($shift$+b) \%2b$] $\gets$ $0$ \label{alg:overview:tune:l2}
				\EndIf
			\EndFor
		\EndProcedure
	\end{algorithmic}
	}
\end{algorithm}

\begin{algorithm}[t]
	\caption{Hash} \label{alg:overview:hash}
	{ \scriptsize
	\begin{algorithmic}[1]
		\Require
			\Statex \textbf{id}: id of record whose hash is being computed
		\Ensure
			\Statex Hash value is returned
		\Procedure{Hash}{id}
			\State \textbf{return} \Call{Z-Value}{$C$[id]} \Comment{\parbox[t]{.375\linewidth}}{Apply $h_{2}$}
		\EndProcedure
	\end{algorithmic}
	}
\end{algorithm}

Algorithms~\ref{alg:overview:initialize}--\ref{alg:overview:hash} present our approach 
	for computing the outcome of \tunableLSH and 
	for incrementally tuning the LSH function every time a query is executed.
Note that we have two design considerations:
\begin{inparaenum}[(i)]
	\item tuning should take constant-time, otherwise, there is no point in using a function,
	\item the memory overhead should be low because it would be desirable to maximize the allocation of memory to core database functionality.
\end{inparaenum}
Consequently,
	instead of relying on record utilization vectors, 
	the algorithm computes and incrementally maintains record utilization counters (cf., Algorithm~\ref{alg:overview:initialize})
	that are much easier to maintain and that have a much smaller memory footprint due to the fact that $b \ll k$.
Then, whenever there is a need to compute the outcome of the LSH function for a given record,
	the \textsc{Hash} procedure is called with the \emph{id} of the record,
	which in turn relies on $h_{2}$ to compute the hash value (cf., Algorithm~\ref{alg:overview:hash}).

The \textsc{Tune} procedure looks at the next query access vector,
	and updates $f$ (line~\ref{alg:overview:tune:la}), which will be discussed in more detail in Section~\ref{sec:details:adaptive}.
Then it	computes positions of records that have been accessed by that query (line~\ref{alg:overview:tune:lb}), and
	increments the corresponding entries in the utilization counters of those records that have been accessed (line~\ref{alg:overview:tune:l1}).
To determine which entry to increment, 
	the algorithm relies on $h_{1}$, hence, $f(t)$ (cf., Def.~\ref{def:tunable-lsh}) and a shifting scheme.
In line~\ref{alg:overview:tune:l2}, old entries in record utilization counters are reset based on an approach that we discuss in Section~\ref{sec:details:reset}.
In that section we also discuss the shifting scheme.

\section{Details of Tunable-LSH and Optimizations}
\label{sec:details}
This section is structured as follows:
Section~\ref{sec:details:properties} shows that \tunableLSH has the properties of a locality-sensitive hashing scheme.
Section~\ref{sec:details:adaptive} describes our approach for tuning $f$ based on the most recent query access patterns, and
Section~\ref{sec:details:reset} explains how old bits are removed from record utilization counters.

\subsection{Properties of Tunable-LSH}
\label{sec:details:properties}

In this part, we discuss the locality-sensitive properties of $h_{1}$ and $h_{2}$, and
	demonstrate that $h_{2} \circ h_{1}$ can be used for clustering the records.
First, we show the relationship between record utilization vectors and
	the record utilization counters that are obtained by applying $h_{1}$.

\begin{theorem}[Distance Bounds]
\label{thm:distance-bounds}

Given a pair of record utilization vectors $\vec{r}_{1}$ and $\vec{r}_{2}$ with size $k$,
	let $\vec{c}_{1}$ and $\vec{c}_{2}$ denote two record utilization counters with size $b$
	such that $\vec{c}_{1} = h_{1}(\vec{r}_{1})$ and $\vec{c}_{2} = h_{1}(\vec{r}_{2})$ (cf., Def.~\ref{def:tunable-lsh}).
Furthermore, let $c_{1}[i]$ and $c_{2}[i]$ denote the $i^{\text{th}}$ entry in $\vec{c}_{1}$ and $\vec{c}_{2}$, respectively.
Then,
\begin{align}
	\delta( \vec{r_{1}}, \vec{r_{2}} ) & \geq \sum\limits_{i=0}^{b-1} \: \bigl\lvert c_{1}[i] - c_{2}[i] \bigr\rvert \label{eqn:lower-bound}
	%%%\\
	%%%\delta( \vec{r_{1}}, \vec{r_{2}} ) & \leq 
	%%%\sum\limits_{i=0}^{b-1} 
	%%%\: \text{min} 
	%%%	\left[
	%%%		\begin{array}{l}
	%%%			c_{1}[i] + c_{2}[i] \\ 
	%%%			2 \lceil \frac{k}{b} \rceil- c_{1}[i] - c_{2}[i] 
	%%%		\end{array}
	%%%	\right] \label{eqn:upper-bound}
\end{align}
where $\delta(\vec{r_{1}}, \vec{r_{2}})$ represents the Hamming distance between $\vec{r}_{1}$ and $\vec{r}_{2}$.
\end{theorem}

\begin{proofSketch}
We prove Thm.~\ref{thm:distance-bounds} by induction on $b$.

\noindent \textbf{Base case}:
Thm.~\ref{thm:distance-bounds} holds when $b=1$.
%%%\begin{itemize}
%%%\item 
%%%\textbf{Eqn~\ref{eqn:lower-bound}}:
According to Def.~\ref{def:tunable-lsh}, when $b=1$, $c_{1}[0]$ and $c_{2}[0]$ correspond to 
	the total number of $1$-bits in $\vec{r}_{1}$ and $\vec{r}_{2}$, respectively.
Note that the Hamming distance between $\vec{r}_{1}$ and $\vec{r}_{2}$ will be smallest 
	if and only if these two record utilization vectors are aligned on as many $1$-bits as possible.
In that case, they will differ in only $\bigl\lvert c_{1}[0] - c_{2}[0] \bigr\rvert$ bits,
	which corresponds to their Hamming distance.
Consequently, Eqn.~\ref{eqn:lower-bound} holds for $b=1$.
%%%\item
%%%\textbf{Eqn~\ref{eqn:upper-bound}}:
%%%The Hamming distance between $\vec{r}_{1}$ and $\vec{r}_{2}$ will be the largest 
%%%	if and only if these two vectors are aligned on as few ``1" bits as possible.
%%%Recall that according to Def.~\ref{def:tunable-lsh} and based on how $f$ has been defined,
%%%	a digit in a record utilization counter can keep track of at most $\lceil \frac{k}{b} \rceil$ bits
%%%	of a record utilization vector.
%%%Therefore, there are two cases to consider.
%%%If $c_{1}[0] + c_{2}[0] \leq \lceil \frac{k}{b} \rceil$, 
%%%	then the vectors can be placed such that their ``1" bits do not overlap at all, 
%%%	hence, their Hamming distance becomes $c_{1}[0] + c_{2}[0]$.
%%%However, if $c_{1}[0] + c_{2}[0] > \lceil \frac{k}{b} \rceil$, 
%%%	according to the Pigeonhole principle~\cite{},
%%%	$c_{1}[0] + c_{2}[0] - \lceil \frac{k}{b} \rceil$ ``1" bits need to overlap.
%%%In that case, the Hamming distance becomes $2 \lceil \frac{k}{b} \rceil- (c_{1}[0] + c_{2}[0])$.
%%%Consequently, Eqn~\ref{eqn:upper-bound} holds for $b=1$.
%%%\end{itemize}

\noindent \textbf{Inductive step}:
We show that if Eqn.~\ref{eqn:lower-bound} holds for $b \leq \alpha$,
	where $\alpha$ is a natural number greater than or equal to $1$,
	then it must also hold for $b = \alpha + 1$.
Let $\Pi_{f}(\vec{r}, g)$ denote a record utilization vector
	$r' = (r'[0], \ldots, r'[k-1])$ such that
	for all $i \in \{0, \ldots, k-1 \}$,
	$r'[i] = r[i]$ holds 
	if $f(i) = g$, and
	$r'[i] = 0$ otherwise.
Then,
\begin{align}
	\delta (\vec{r}_{1}, \vec{r}_{2}) = \sum \limits_{g=0}^{b-1} \delta (\Pi_{f}(\vec{r}_{1}, g), \Pi_{f}(\vec{r}_{2}, g)). \label{eqn:additive-property}
\end{align}
That is, the Hamming distance between any two record utilization vectors is
	the summation of their individual Hamming distances within each group of bits 
	that share the same hash value with respect to $f$.
This property holds because $f$ is a (total) function, and $\Pi_{f}$ masks all the irrelevant bits.
As an abbreviation, let $\delta_{g} = \delta (\Pi_{f}(\vec{r}_{1}, g), \Pi_{f}(\vec{r}_{2}, g))$.
Then, due to the same reasoning as in the base case,
	for $g=\alpha$,	the following equation holds:
\begin{align}
	\delta_{\alpha}( \vec{r_{1}}, \vec{r_{2}} ) & \geq \bigl\lvert c_{1}[\alpha] - c_{2}[\alpha] \bigr\rvert \label{eqn:inductive:lower-bound} 
	%%%\\ 	
	%%%\delta_{\alpha}( \vec{r_{1}}, \vec{r_{2}} ) & \leq 
	%%%\text{min} 
	%%%	\left[
	%%%		\begin{array}{l}
	%%%			c_{1}[\alpha] + c_{2}[\alpha] \\
	%%%			2\lambda[\alpha] - c_{1}[\alpha] - c_{2}[\alpha]
	%%%		\end{array}
	%%%	\right] . \label{eqn:inductive:upper-bound}
\end{align}
Consequently, using the additive property in Eqn.~\ref{eqn:additive-property}, 
	it can be shown that Eqn~\ref{eqn:lower-bound} holds also for $b = \alpha+1$.
Thus, by induction, Thm.~\ref{thm:distance-bounds} holds. \hfill $\blacksquare$
\end{proofSketch}

Thm.~\ref{thm:distance-bounds} suggests that the Hamming distance between 
	any two record utilization vectors $\vec{r_{1}}$ and $\vec{r_{2}}$ can be approximated
	using record utilization counters $\vec{c_{1}} = h_{1} (\vec{r_{1}})$ and $\vec{c_{2}} = h_{1} (\vec{r_{2}})$ because
		Eqn.~\ref{eqn:lower-bound} provides a lower bound 	on $\delta (\vec{r_{1}}, \vec{r_{2}})$.
In fact, the right-hand side of Eqn.~\ref{eqn:lower-bound} is equal to 
	the Manhattan distance~\cite{Krause1986} between $\vec{c_{1}}$ and $\vec{c_{2}}$ in $\intRangeCart{0}{\lceil \frac{k}{b} \rceil}{b}$,
	and since $\delta(\vec{r_{1}}, \vec{r_{2}})$ is equal to 
	the Manhattan distance between $\vec{r_{1}}$ and $\vec{r_{2}}$ in $\intRangeCart{0}{1}{k}$,
	it is easy to see that $h_{1}$ is a transformation that approximates Manhattan distances.
The following corollary captures this property.

\begin{corollary}[Distance Approximation]
Given a pair of record utilization vectors $\vec{r}_{1}$ and $\vec{r}_{2}$ with size $k$,
	let $\vec{c_{1}}$ and $\vec{c_{2}}$ denote two points 
	in the coordinate system $\intRangeCart{0}{\lceil \frac{k}{b} \rceil}{b}$ such that
	$\vec{c_{1}} = h_{1} (\vec{r}_{1})$ and $\vec{c_{2}} = h_{1} (\vec{r}_{2})$ (cf., Def.~\ref{def:tunable-lsh}).
Let $\delta^{M} (\vec{r_{1}}, \vec{r_{2}})$ denote the Manhattan distance	
	between $\vec{r_{1}}$ and $\vec{r_{2}}$, and
let $\delta^{M} (\vec{c_{1}}, \vec{c_{2}})$ denote the Manhattan distance
	between $\vec{c_{1}}$ and $\vec{c_{2}}$.
Then, the following holds:
	\begin{align}
		\delta (\vec{r}_{1}, \vec{r}_{2}) = \delta^{M} (\vec{r_{1}}, \vec{r_{2}}) &\geq \delta^{M} (\vec{c_{1}}, \vec{c_{2}}) \label{eqn:manhattan:lower}
	\end{align}
\end{corollary}

\begin{proofSketch}
	Hamming distance in $\intRangeCart{0}{1}{k}$ is a special case of Manhattan distance.
	Furthermore, by definition~\cite{Krause1986}, the right hand side of Eqn.~\ref{eqn:lower-bound} 
		equals the Manhattan distance $\delta^{M} (\vec{c_{1}}, \vec{c_{2}})$; therefore,
		Eqn.~\ref{eqn:manhattan:lower} holds. \hfill $\blacksquare$
\end{proofSketch}

Next, we demonstrate that $h_{1}$ is a locality-sensitive transformation~\cite{IndykSTOC98,GionisVLDB1999}.
In particular, we use the definition of locality-sensitiveness by Tao et al.~\cite{TaoTODS2010}, and show that the probability that
	two record utilization vectors $\vec{r_{1}}$ and $\vec{r_{2}}$ are transformed into
	``near\-by" record utilization counters $\vec{c_{1}}$ and $\vec{c_{2}}$
	increases as the (Manhattan) distance between $r_{1}$ and $r_{2}$ decreases.

\begin{theorem}[Good Approximation]
	\label{thm:probabilities}
	Gi\-ven a pair of re\-cord utilization vectors $\vec{r_{1}}$ and $\vec{r_{2}}$ with size $k$,
		let $\vec{c_{1}}$ and $\vec{c_{2}}$ denote two points in the coordinate system
		$\intRangeCart{0}{\lceil \frac{k}{b} \rceil}{b}$ such that 
		$\vec{c_{1}} = h_{1} (\vec{r_{1}})$,
		$\vec{c_{2}} = h_{1} (\vec{r_{2}})$ and
		$b=1$ (cf., Def.~\ref{def:tunable-lsh}).
	Let $\delta^{M} (\vec{r_{1}}, \vec{r_{2}})$ denote the Manhattan distance between $\vec{r_{1}}$ and $\vec{r_{2}}$, and
		let $\delta^{M} (\vec{c_{1}}, \vec{c_{2}})$ denote the Manhattan distance between 
		$\vec{c_{1}}$ and $\vec{c_{2}}$.
	Furthermore, let $\textsc{Pr}_{\delta^{M} \leq \Theta}(x)$ be a shorthand for 
	\begin{align*}
		\textsc{Pr} \Big( \delta^{M}(\vec{c_{1}}, \vec{c_{2}}) \leq \Theta \; \Big| \; \delta^{M}(\vec{r_{1}}, \vec{r_{2}})=x \Big).	
	\end{align*}
	Then,
	\begin{align}
		\textsc{Pr}_{\delta^{M} \leq \Theta} (x) = 
		\frac {\mathlarger{\sum \limits_{i = \lceil \frac{x - \Theta}{2} \rceil}^{\lfloor \frac{x + \Theta}{2} \rfloor}} \binom{x}{i}}{2^{x}}
		\label{eqn:probabilities}
	\end{align}
	where $\Theta, x \in \intRange{0}{\lceil \frac{k}{b} \rceil}$ such that $\Theta < x$.
\end{theorem}

\begin{proofSketch}
	If the Hamming/Manhattan distance between $\vec{r_{1}}$ and $\vec{r_{2}}$ is $x$,
	 then it means that these two vectors will differ in exactly $x$ bits, 
	as shown below.
	\begin{align*}
		\vec{\mathbf{r_{1}}} &: \; \Box\Box\Box \, \overbrace{\mathbf{111 \ldots 1}}^{a} \, \mathbf{0 \cdots 000} \, \Box\Box\Box \\
		\vec{\mathbf{r_{2}}} &: \; \Box\Box\Box \, \mathbf{000 \cdots 0} \, \underbrace{\mathbf{1 \ldots 111}}_{x-a} \, \Box\Box\Box
	\end{align*}
	Furthermore, if $\vec{r_{1}}$ has $\Delta + a$ bits set to $1$, then
		$\vec{r_{2}}$ must have $\Delta + (x-a)$ bits set to $1$,
		where $\Delta$ denotes the number of matching $1$-bits between $\vec{r_{1}}$ and $\vec{r_{2}}$ and
		$a \in \{ 0, \ldots, x \}$.
	Note that when $b=1$, the Manhattan distance between $\vec{c_{1}}$ and $\vec{c_{2}}$ is 
		equal to the difference in the number of $1$-bits that $\vec{r_{1}}$ and $\vec{r_{2}}$ have.
	Hence,
	\begin{align*}
		\delta^{M} (\vec{c_{1}}, \vec{c_{2}}) &= \big| \Delta + x - a - (\Delta + a) \big| \\
																				&= \big| x - 2a \big|.
	\end{align*}
	It is easy to see that there are $(x+1)$ different configurations:
	\begin{align*}
		a &= 0 & \Rightarrow & & &\delta^{M}(\vec{c_{1}}, \vec{c_{2}}) = x \\
		a &= 1 & \Rightarrow & & &\delta^{M}(\vec{c_{1}}, \vec{c_{2}}) = x\!-\!2 \\
		& \; \; \vdots & & & & \\
		a &= x\!-\!1 & \Rightarrow & & &\delta^{M}(\vec{c_{1}}, \vec{c_{2}}) = x\!-\!2 \\
		a &= x & \Rightarrow & & &\delta^{M}(\vec{c_{1}}, \vec{c_{2}}) = x.
	\end{align*}
	Only when $a = \big\{ \lceil \frac{x-\Theta}{2} \rceil, \ldots, \lfloor \frac{x+\Theta}{2} \rfloor \big\}$,
		will $\delta^{M} (\vec{c_{1}}, \vec{c_{2}}) \leq \Theta$ be satisfied.
	For each satisfying value of $a$,
		the non-matching bits in $\vec{r_{1}}$ and $\vec{r_{2}}$ 
		can be combined in $\binom{x}{a}$ possible ways.
	Therefore, there are a total of
	\begin{align*}
		\sum \limits_{i = \lceil \frac{x-\Theta}{2} \rceil}^{\lfloor \frac{x+\Theta}{2} \rfloor} \binom{x}{i}
	\end{align*}
	combinations such that $\delta^{M} (\vec{c_{1}}, \vec{c_{2}}) \leq \Theta$.
	Since there are $2^{x}$ possible combinations in total, the posterior probability in Thm.~\ref{thm:probabilities} holds. 
	
	\hfill $\blacksquare$
\end{proofSketch}

Using Thm.~\ref{thm:probabilities}, it is possible to show that when $b=1$, for all $\Theta < x$ where $\Theta, \: x \in \intRange{0}{(k-2)}$, the following holds:
\begin{align}
	Pr_{\delta^{M} \leq \Theta}(x) > Pr_{\delta^{M} \leq \Theta}(x\!+\!2). \label{eqn:lsh-property}
\end{align}
Therefore, $h_{1}$ is locality-sensitive for $b=1$.
Due to space limitations, we omit the proof of Eqn.~\ref{eqn:lsh-property}, but in a nutshell,
	the proof follows from the fact that going from $x$ to $x+2$, 
	the denominator in Eqn.~\ref{eqn:probabilities} always increases by a factor of $4$,
	whereas the numerator increases by a factor that is strictly less than $4$.

\begin{figure}[t]
	\begin{center}
		\includegraphics[width=0.95\linewidth]{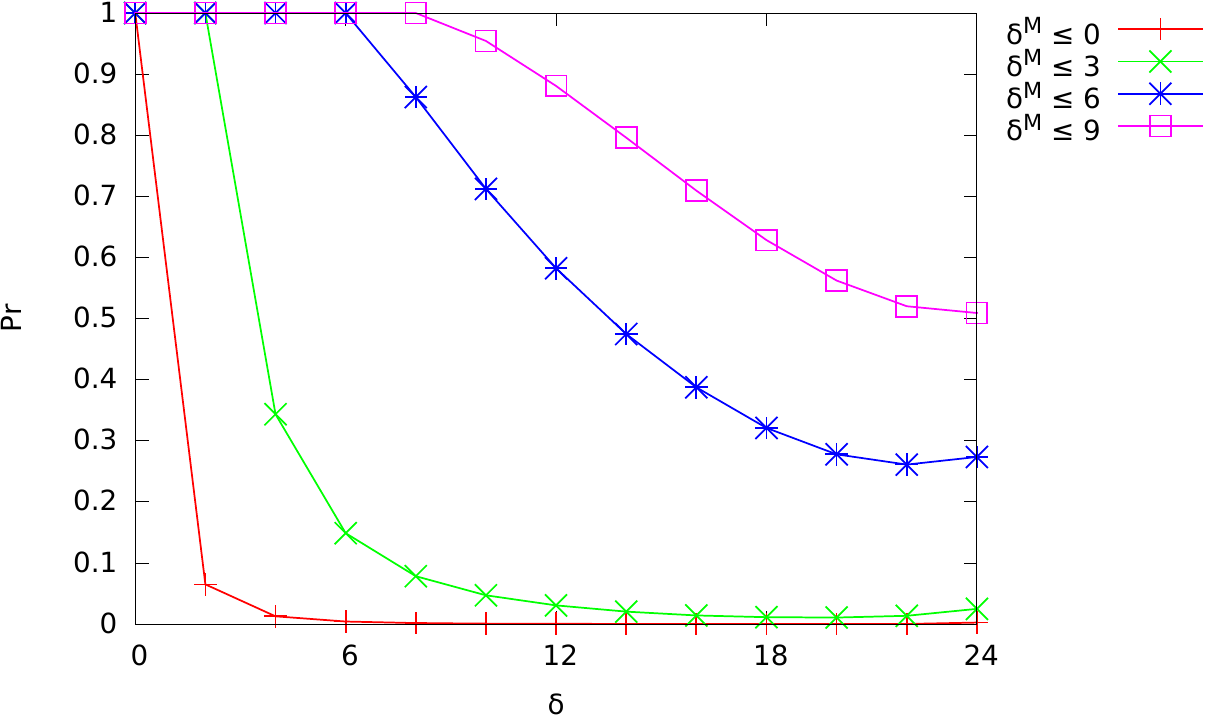}
	\end{center}
	\caption{$\textsc{Pr}_{\delta^{M} \leq \Theta}$ for $k=24$ and $b=6$}
	\label{fig:multi_b_probability}
\end{figure}

Generalizing Thm.~\ref{thm:probabilities} and Eqn.~\ref{eqn:lsh-property} to cases where $b \geq 2$ is more complicated.
However, our empirical analyses across multiple values of $k$ and $b$ demonstrate that
\begin{align*}
	Pr_{\delta^{M} \leq \Theta} (x) \gg Pr_{\delta^{M} \leq \Theta} (y)
\end{align*}
holds when $y \gg x$.
For example, Fig.~\ref{fig:multi_b_probability} shows $Pr_{\delta^{M} \leq \Theta} (x)$ when $k=24$ and $b=6$.
Fig.~\ref{fig:multi_b_probability}, along with our empirical evaluations, verify that $h_{1}$ is locality sensitive.
Thus, combined with a space-filling curve,
	it can be used to approximate the clustering problem.

\subsection{Achieving and Maintaining Tighter Bounds on Tunable-LSH}
\label{sec:details:adaptive}

Next, we demonstrate how it is possible to reduce the approximation error of $h_{1}$.
We first define \emph{load factor} of an entry of a record utilization counter.

\begin{definition}[Load Factor]
	Given a record utilization counter $\vec{c} = (c[0], \ldots, c[b\!-\!1])$ with size $b$,
		the \emph{load factor} of the $i^{th}$ entry is $c[i]$.
\end{definition}

\begin{theorem}[Effects of Grouping]
	\label{thm:error-wrt-load}
	Given two record utilization vectors $\vec{r_{1}}$ and $\vec{r_{2}}$ with size $k$,
		let $\vec{c_{1}}$ and $\vec{c_{2}}$ denote two record utilization counters with size $b=1$
		such that $\vec{c_{1}} = h_{1} (\vec{r_{1}})$ and $\vec{c_{2}} = h_{1} (\vec{r_{2}})$.
	Then,
	\begin{align}
		\textsc{Pr} \left(
			\begin{array}{l}
				\delta^{M}(\vec{c_{1}}, \vec{c_{2}}) \\
				\; \; = \delta^{M}(\vec{r_{1}}, \vec{r_{2}}) 
			\end{array}
				\: \Bigg| \:
			\begin{array}{l}
				c_{1}[0] = l_{1} \; \textsc{and} \\
				c_{2}[0] = l_{2} 
			\end{array}
			\right)
				= \gamma \label{eqn:error-wrt-load:part1}
	\end{align}
	where
	\begin{align}
		\gamma = \frac { \binom{l_{\text{max}}}{l_{\text{min}}} \binom{k}{l_{\text{max}}} } { \binom{k}{l_{\text{max}}} \binom{k}{l_{\text{min}}} } \label{eqn:error-wrt-load:part2}
	\end{align}
	and
	\begin{align*}
		l_{\text{max}} &= max (l_{1}, l_{2}) \\
		l_{\text{min}} &= min (l_{1}, l_{2}).
	\end{align*}
\end{theorem}

\begin{proofSketch}
	Let $\vec{r}_{\text{max}}$ denote the record utilization vector with the most number of $1$-bits among $\vec{r_{1}}$ and $\vec{r_{2}}$, and
		let $\vec{r}_{\text{min}}$ denote the vector with the least number of $1$-bits.
	When $b=1$, $\delta^{M} (\vec{c_{1}}, \vec{c_{2}}) = \delta (\vec{r_{1}}, \vec{r_{2}})$ holds if and only if
		the number of $1$-bits on which $\vec{r_{1}}$ and $\vec{r_{2}}$ are aligned is $l_{\text{min}}$
		because in that case, both $\delta^{M} (\vec{c_{1}}, \vec{c_{2}})$ and $\delta (\vec{r_{1}}, \vec{r_{2}})$ are equal to
		$l_{\text{max}} - l_{\text{min}}$ (note that $\delta^{M} (\vec{c_{1}}, \vec{c_{2}})$ is always equal to $l_{\text{max}} - l_{\text{min}}$).
	Assuming that the positions of $1$-bits in $\vec{r}_{\text{max}}$ are fixed,
		there are $\binom{l_{\text{max}}}{l_{\text{min}}}$ possible ways of arranging the $1$-bits of $\vec{r}_{\text{min}}$ such that
		$\delta (\vec{r}_{1}, \vec{r}_{2}) = l_{\text{max}} - l_{\text{min}}$.
	Since the $1$-bits of $\vec{r}_{\text{max}}$ can be arranged in $\binom{k}{l_{\text{max}}}$ different ways, there are
		$\binom{l_{\text{max}}}{l_{\text{min}}} \binom{k}{l_{\text{max}}}$ combinations such that
		$\delta^{M} (\vec{c}_{1}, \vec{c}_{2}) = \delta (\vec{r}_{1} , \vec{r}_{2})$.
	Note that in total, the bits of $\vec{r_{1}}$ and $\vec{r_{2}}$ can be arranged in
		$\binom{k}{l_{\text{max}}} \binom{k}{l_{\text{min}}}$ possible ways; 
		therefore, Eqns.~\ref{eqn:error-wrt-load:part1} and~\ref{eqn:error-wrt-load:part2} describe the posterior probability that
		$\delta^{M} (\vec{c_{1}}, \vec{c_{2}}) = \delta (\vec{r_{1}}, \vec{r_{2}})$,
		given	$c_{1}[0] = l_{1}$ and $c_{2}[0] = l_{2}$. \hfill $\blacksquare$
\end{proofSketch}

\begin{figure}[t]
	\begin{center}
		\includegraphics[width=0.95\linewidth]{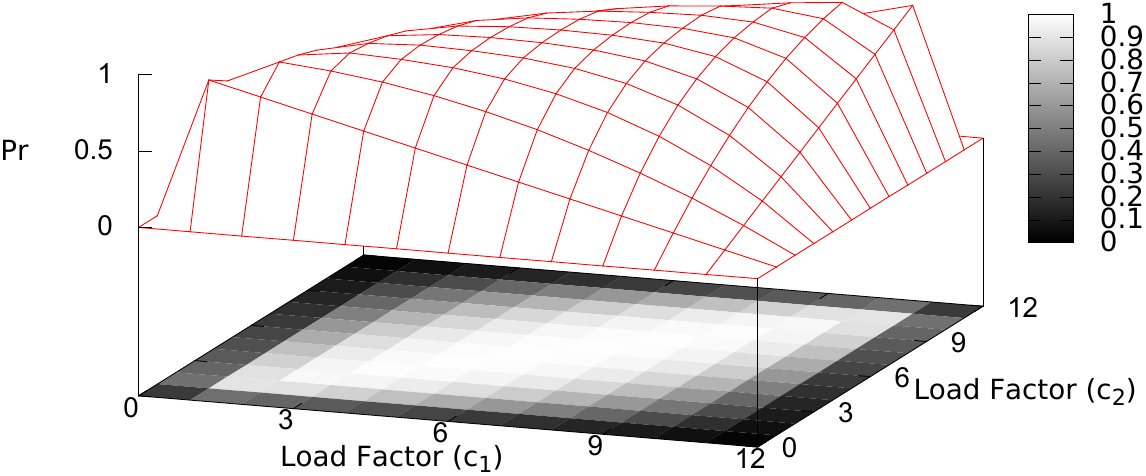}
	\end{center}
	\caption{$\textsc{Pr}(\delta^{M} \neq \delta)$ for $k=12$, $b=1$ and across varying load factors}
	\label{fig:error_wrt_load}
\end{figure}

According to Eqns.~\ref{eqn:error-wrt-load:part1} and~\ref{eqn:error-wrt-load:part2} in Thm.~\ref{thm:error-wrt-load},
	the probability that $\delta^{M} (\vec{c_{1}},\-\vec{c_{2}})$ is an approximation of $\delta (\vec{r_{1}}, \vec{r_{2}})$,
	but that it is not exactly equal to $\delta (\vec{r_{1}}, \vec{r_{2}})$ is lower
	for load factors that are close or equal to zero and
	likewise for load factors that are close or equal to $\lceil \frac{k}{b} \rceil$ (cf., Fig.~\ref{fig:error_wrt_load}).
This property suggests that by carefully choosing $f$,
	it is possible to achieve even tighter error bounds for $h_{1}$.

Contrast the matrices in Fig~\ref{fig:matrix-rep:rows} and Fig~\ref{fig:matrix-rep:columns},
	which contain the same query access vectors, but the columns are grouped in two different ways\footnote{Groups are separated by vertical dashed lines.}:
	\begin{inparaenum}[(i)]
		\item in Fig.~\ref{fig:matrix-rep:rows}, the grouping is based on the original sequence of execution, and
		\item in Fig.~\ref{fig:matrix-rep:columns}, queries with similar access patterns are grouped together.
	\end{inparaenum}
Fig.~\ref{fig:rows:transformed} and Fig.~\ref{fig:rc:transformed} represent 
	the corresponding record utilization counters for 
	the record utilization vectors in the matrices in
	Fig.~\ref{fig:matrix-rep:rows} and Fig.~\ref{fig:matrix-rep:columns}, respectively.
Take $\record{r_{3}}$ and $\record{r_{5}}$, for instance.
Their actual Hamming distance with respect to $q_{0}$--$q_{7}$ is $8$.
Now consider the transformed matrices.
According to Fig.~\ref{fig:rows:transformed}, the Hamming distance lower bound is $0$,
	whereas according to Fig.~\ref{fig:rc:transformed}, it is $8$.
Clearly, the bounds in the second representation are closer to the original.
The reason is as follows.
Even though $\record{r_{3}}$ and $\record{r_{5}}$ differ on all the bits for $q_{0}$--$q_{7}$,
	when the bits are grouped as in Fig.~\ref{fig:matrix-rep:rows},
	the counts alone cannot distinguish the two bit vectors.
In contrast, if the counts are computed based on the grouping in Fig.~\ref{fig:matrix-rep:columns}
	(which clearly places the $1$-bits in separate groups),
	the counts indicate that the two bit vectors are indeed different.

The observations above are in accordance with Thm.~\ref{thm:error-wrt-load}.
Consequently, we make the following optimization.
Instead of randomly choosing a hash function,
	we construct $f$ such that 
	it maps queries with similar access vectors (i.e., columns in the matrix)
	to the same hash value.
This way, it is possible to obtain record utilization counters with entries that have
	either very high or very low load factors (cf., Def.~\ref{def:tunable-lsh}),
	thus, decreasing the probability of error (cf., Thm.~\ref{thm:error-wrt-load}).

We develop a technique to efficiently determine 
	groups of queries with similar access patterns and 
	to adaptively maintain these groups as the access patterns change.
Our approach consists of two parts:
\begin{inparaenum}[(i)]
	\item to approximate the similarity between any two queries, we rely on the \textsc{Min-Hash} scheme~\cite{BroderSEQUENCES1997}, and
	\item to adaptively group similar queries, we develop an incremental version of a multidimensional scaling (MDS) algorithm~\cite{MorrisonINFVIS2003}.
\end{inparaenum}

\textsc{Min-Hash} offers a quick and efficient way of approximating the similarity,
	(more specifically, the Jaccard similarity~\cite{JaccardNP1912}), between two sets of integers.
Therefore, to use it, 
	the query access vectors in our conceptualization need to be translated into a set of positional identifiers
	that correspond to the records for which 
	the bits in the vector are set to $1$.\footnote{In practice, this translation never takes place because the system maintains positional vectors to begin with.}	
For example, according to Fig.~\ref{fig:matrix-rep:original}, 
	$\vec{q_{1}}$ should be represented with the set $\{0, 5, 6\}$ 
	because $r_{0}$, $r_{5}$ and $r_{6}$ are the only records for which the bits are set to $1$.
Note that, we do not need to store the original query access vectors at all.
In fact, after the access patterns over a query are determined, 
	we compute and store only its \textsc{Min-Hash} value.
This is important for keeping the memory overhead of our algorithm low.

\begin{table}[t]
	\dataStructuresTable
	\caption{Data structures referenced in algorithms}
	\label{tab:data-structures}
\end{table}

Queries with similar access patterns are grouped together
	using a multidimensional scaling (MDS) algorithm~\cite{KruskalPsychometrika1964} that was
	originally developed for data visualization, and
	has recently been used for clustering~\cite{BislimovskaEDBT2015}.
Given a set of points and a distance function, 
	MDS assigns coordinates to points such that their original distances are preserved as much as possible.
In one efficient implementation~\cite{MorrisonINFVIS2003},
	each point is initially assigned a random set of coordinates, 
	but these coordinates are adjusted iteratively based on a spring-force analogy.
That is, it is assumed that points exert a force on each other that is proportional
	to the difference between their actual and observed distances,
	where the latter refers to the distance that is computed from
	the algorithm-assigned coordinates.
These forces are used for computing the current velocity ($V$ in Table~\ref{tab:data-structures})
	and the approximated coordinates of a point ($X$ in Table~\ref{tab:data-structures}).
The intuition is that, after successive iterations,
	the system will reach equilibrium, at which point,
	the approximated coordinates can be reported.
Since computing all pairwise distances can be prohibitively expensive,
	the algorithm relies on a combination of sampling ($S[]$ in Table~\ref{tab:data-structures}) 
	and maintaining for each point, a list of its nearest neighbours ($N[]$ in Table~\ref{tab:data-structures})---only these
	distances are used in computing the net force acting on a point.
Then, the nearest neighbours are updated in each iteration
	by removing the most distant neighbour of a point and replacing it with a new
	point from the random sample if the distance between the point and the random sample
	is smaller than the distance between the point and its most distant neighbour.

\begin{algorithm}[t]
	\caption{Reconfigure-F} \label{alg:online}
	{ \scriptsize
	\begin{algorithmic}[1]
		\Require
			\Statex $\vec{\mathbf{q_{t}}}$: query access vector produced at time $t$
		\Ensure
			\Statex Coordinates of MDS points are updated, which are used in determining the outcome of $f$
		\Procedure{Reconfigure-F}{$\vec{q_{t}}$}
			\State{$\text{pos} \leftarrow$ $( \text{begin} + \text{size} ) \: \% \: k$} \label{alg:online:l1}
			\State{$S[\text{pos}]\text{.clear()}$} \label{alg:online:l2}
			\State{$N[\text{pos}]\text{.clear()}$} \label{alg:online:l3}
			\State{$X[\text{pos}] \leftarrow -0.5 + rand() \: / \: \textsc{rand-max}$} \label{alg:online:l4}
			\State{$V[\text{pos}] \leftarrow 0$} \label{alg:online:l5}
			\State{$H[\text{pos}] \leftarrow \Call{Min-Hash}{\vec{q_{t}}}$} \label{alg:online:l6}
			\If{$\text{size} < k$}
				\State{size \verb!+=! $1$}
			\Else
				\State{$\text{begin} = (\text{begin} + 1) \: \% \: k$}
			\EndIf
			\For{$i \leftarrow 0, i < \text{size}, i\texttt{++}$} \label{alg:online:l7}
				\State{$x \leftarrow (\text{begin}+i) \: \% \: k$}
				\State \Call{Update-S-and-N}{$x$} \label{alg:online:l8}
				\State \Call{Update-Velocity}{$x$} \label{alg:online:l9}
			\EndFor
			\For{$i \leftarrow 0, i < \text{size}, i\texttt{++}$}
				\State{$x \leftarrow (\text{begin}+i) \: \% \: k$} 				
				\State \Call{Update-Coordinates}{$x$} \label{alg:online:l10}
			\EndFor \label{alg:online:l11}
		\EndProcedure
	\end{algorithmic}
	}
\end{algorithm}

This algorithm cannot be used directly for our purposes because it is not incremental.
Therefore, we propose a revised MDS algorithm that incorporates the following modifications:
\begin{enumerate}
	\item In our case, each point in the algorithm represents a query access vector.
			However, since we are not interested in visualizing these points, but rather clustering them,
				we configure the algorithm to place these points along a single dimension.
			Then, by dividing the coordinate space into consecutive regions, 
				we are able to determine similar query access vectors.
	\item Instead of computing the coordinates of all of the points at once,
				our version makes incremental adjustments to the coordinates 
				every time reconfiguration is needed.
\end{enumerate}

The revised algorithm is given in Algorithm~\ref{alg:online}.
First, the algorithm decides which MDS point to assign 
	to the new query access vector $\vec{q_{t}}$ (line~\ref{alg:online:l1}).
It clears the array and the heap data structures containing, respectively,
	\begin{inparaenum}[(i)]
		\item the randomly sampled, and
		\item the neighbouring set of points (lines~\ref{alg:online:l2}--\ref{alg:online:l3}).
	\end{inparaenum}
Furthermore, it assigns a random coordinate to the point within the interval $[-0.5, 0.5]$ (line~\ref{alg:online:l4}), and
	resets its velocity to $0$ (line~\ref{alg:online:l5}).
Next, it computes the \textsc{Min-Hash} value of $\vec{q_{t}}$ and stores it in $H[\text{pos}]$ (line~\ref{alg:online:l6}).
Then, it makes two passes over all the points in the system (lines~\ref{alg:online:l7}--\ref{alg:online:l11}), while 
	first updating their sample and neighbouring lists (line~\ref{alg:online:l8}),
	computing the net forces acting on them based on the \textsc{Min-Hash} distances and
	updating their velocities (line~\ref{alg:online:l9}); 
	and then updating their coordinates (line~\ref{alg:online:l10}).

The procedures used in the last part are implemented in a similar way as the original algorithm~\cite{MorrisonINFVIS2003}; that is, 
	in line~\ref{alg:online:l8}, the sampled points are updated, 
	in line~\ref{alg:online:l9}, the velocities assigned to the MDS points are updated, and
	in line~\ref{alg:online:l10}, the coordinates of the MDS points are updated based on these updated velocities.
However, our implementation of the \textsc{Update-Velocity} procedure (line~\ref{alg:online:l9}) is slightly different than the original.
In particular, in updating the velocities, we use a decay function so that
	the algorithm forgets ``old" forces that might have originated from
	the elements in $S[]$ and $N[]$ that have been assigned to new
	query access vectors in the meantime.
Note that unless one keeps track of the history of all the forces
	that have acted on every point in the system, 
	there is no other way of ``undoing" or ``forgetting" these ``old" forces.

\begin{algorithm}[t]
	\caption{Hash Function $f$} \label{alg:hash-function-f}
	{ \scriptsize
	\begin{algorithmic}[1]
		\Require
			\Statex $\mathbf{t}$: sequence number of a query access vector
		\Ensure
			\Statex $f(t)$ is computed and returned
		\Procedure{f}{$t$}
			\State pos $\gets$ $t \: \% \: k$
			\State $(\text{lo}, \text{hi}) \leftarrow$ \Call {group-bounds}{$X[$pos$]$}
			\State $\text{coid} \leftarrow$ \Call {centroid} {$\text{lo}, \text{hi}$}
			\State \Return \Call {hash} {$\text{coid}$} $\% \, b$
		\EndProcedure		
	\end{algorithmic}
	}
\end{algorithm}

Given the sequence number of a query access vector ($t$),
	the outcome of the hash function $f$ is determined
	based on the coordinates of the MDS point that had previously been
	assigned to the query access vector by 
	the \textsc{Reconfigure} procedure.
To this end, the coordinate space is divided into $b$ groups
	containing points with consecutive coordinates such that
	there are at most $\lceil \frac{k}{b} \rceil$ points in each group.
Then, one option is to use the group identifier, which is a number in $\intRange{0}{b-1}$,
	as the outcome of $f$, but
	there is a problem with this na\"ive implementation.
Specifically, we observed that 
	even though the \emph{relative} coordinates of MDS points 
	within the ``same" group may not change significantly 
	across successive calls to the \textsc{Reconfigure} procedure,
	points within a group, as a whole, may shift.
This is an inherent (and in fact, a desirable) property of the incremental algorithm.
However, the problem is that there may be far too many cases 
	where the group identifier of a point changes 
	just because the absolute coordinates of the group have changed, 
	even though the point continues to be part of the ``same" group.
To solve this problem,
	we rely on a method of computing the centroid within a group
	by taking the \textsc{Min-Hash} of the identifiers of points within that group
	such that these centroids rarely change across successive iterations.
Then, we rely on the identifier of the centroid, as opposed to its coordinates,
	to compute the group number, hence, the outcome of $f$.
The pseudocode of this procedure is given in Algorithm~\ref{alg:hash-function-f}.

We make one last observation.
Internally, \textsc{Min-Hash} uses multiple hash functions to approximate the degree to which two sets are similar~\cite{BroderSEQUENCES1997}.
It is also known that increasing the number of internal hash functions used (within \textsc{Min-Hash})
should increase the overall accuracy of the \textsc{Min-Hash} scheme.
However, as unintuitive as it may seem, in our approach, we use only a single hash function within \textsc{Min-Hash},
	yet, we are still able to achieve sufficiently high accuracy.
The reason is as follows.
Recall that Algorithm~\ref{alg:online} relies on 
	multiple pairwise distances to position every point.
Consequently, even though individual pairwise distances may be inaccurate (because we are just using a single hash function within \textsc{Min-Hash}),
	collectively the errors are cancelled out, and points can be positioned accurately on the MDS coordinate space.

\subsection{Resetting Old Entries in Record Utilization Counters}
\label{sec:details:reset}

\begin{figure}[t]
	\centering
	\counterShifting
	\caption{Assuming $b=3$, $\Box$ indicates the allowed locations at each time tick, and $\emptyset$ indicates the counter to be reset.}
	\label{fig:counter-shifting}
\end{figure}

Once the group identifier is computed (cf., Algorithm~\ref{alg:hash-function-f}),
	it should be straightforward to update 
	the record utilization counters (cf., line~\ref{alg:overview:tune:l1} in Algorithm~\ref{alg:overview:tune}).
However, unless we maintain the original query access vectors, 
	we have no way of knowing which counters to decrement
	when a query access vector becomes stale,
	as maintaining these original query access vectors is prohibitively expensive.
Therefore, we develop a more efficient scheme in which
	old values can also be removed from the record utilization counters.

Instead of maintaining $b$ entries in every record utilization counter,
	we maintain twice as many entries ($2b$).
Then, whenever the \textsc{Tune} procedure is called,
	instead of directly using the outcome of $f(t)$ 
	to locate the counters to be incremented, 
	we map $f(t)$ to a location within an ``allowed"
	region of consecutive entries in the record utilization counter 
	(cf., line~\ref{alg:overview:tune:l1} in Algorithm~\ref{alg:overview:tune}).
At every $\lceil \frac{k}{b} \rceil^{\text{th}}$ iteration,
	this allowed region is shifted by one to the right,
	wrapping back to the beginning if necessary.
Consider Fig.~\ref{fig:counter-shifting}.
Assuming that $b=3$ and that at time $t=0$
	the allowed region spans entries from $0$ to $(b-1)$,
	at time $t=\lceil \frac{k}{b} \rceil$, the region will span entries from $1$ to $b$;
	at time $t=k$, the region will span entries from $b$ to $2b-1$; and
	at time $t=\lceil \frac{4k}{b} \rceil$, the region will span entries $0$ and those from $b+1$ to $2b-1$.

Since $f(t)$ produces a value between $0$ and $b-1$ (inclusive), whereas the entries are numbered from $0$ to $2b-1$ (inclusive),
	the \textsc{Reconfigure} procedure in Algorithm~\ref{alg:overview:tune} uses $f(t)$ as follows.
If the outcome of $f(t)$ directly corresponds to a location in the allowed region, then it is used.
Otherwise, the output is incremented by $b$ (cf., line~\ref{alg:overview:tune:l1} in Algorithm~\ref{alg:overview:tune}).
Whenever the allowed region is shifted to the right, it may land on an already incremented entry.
If that is the case, that entry is reset, thereby allowing ``old" values forgotten (cf., line~\ref{alg:overview:tune:l2} in Algorithm~\ref{alg:overview:tune}).
These are shown by $\emptyset$ in Fig.~\ref{fig:counter-shifting}.
This scheme guarantees any query access pattern that is less than $k$ steps old is remembered,
	while any query access pattern that is more than $2k$ old is forgotten.

%%To wrap up, we have introduced \tunableLSH, which is a tunable locality-sensitive hashing scheme.
%%We have demonstrated that the original Hamming distances between record utilization vectors can be approximated using \tunableLSH with tight error bounds.
%%We have also introduced an adaptive clustering step, in which queries are pre-clustered based on their access pattern similarity, 
%%	which not only improves the error bounds of \tunableLSH, 
%%	but also ensures that these tighter error bounds are maintained even when the query access patterns change.
%%These properties of \tunableLSH enable us to use it for clustering records in an RDF data management system, which we discuss next.

%%%The clustering algorithm we propose in this paper is oblivious to 
%%%\emph{how} the last $k$ most representative query access vectors are selected.
%%%Therefore, if more effective replacement schemes are developed (which is orthogonal to our work), they can be easily integrated.
%%%As for the choice of $k$, without exceeding the memory budget that is allocated to our algorithm, 
%%%we would like to keep it as large as possible to achieve better prediction accuracy.

%\section{Clustering using Tunable-LSH}
%\label{sec:physical-clustering}
%\input{sections/caching}

\section{Experimental Evaluation}
\label{sec:evaluation}
In this section, we evaluate \tunableLSH in three sets of experiments.
First, we evaluate it within \emph{chameleon-db}, our prototype RDF data management system~\cite{AlucUW2013}.
Second, we evaluate it within a hashtable implementation since hashtables are used extensively in RDF data management systems.
Finally, we evaluate \tunableLSH in isolation, to understand how it behaves under different types of workloads.
All experiments are performed on a commodity machine with 
AMD Phenom II $\times4$ $955$ $3.20$~GHz processor, 
$16$~GB of main memory and
a hard disk drive with $100$~GB of free disk space.
The operating system is Ubuntu $12.04$ LTS.

\subsection{Tunable-LSH in chameleon-db}

\begin{table}[t]
	{
		\scriptsize
		\begin{tabular}{r | r | r | r | r | r | r | r}
				& \rot{\textbf{CDB [ICDE'15]}}
				& \rot{\textbf{CBD [\tunableLSH]}}			
				& \rot{\textbf{RDF-3x}}
				& \rot{\textbf{VOS [6.1]}}
				& \rot{\textbf{VOS [7.1]}}
				& \rot{\textbf{MonetDB}}
				& \rot{\textbf{4Store}} \\ \hline
				WatDiv $10$M	& $\mathbf{4.7}$	& $19.4$ 		& $18.8$	& $44.0$	& $24.5$	& $17.0$	& $93.0$	\\ \hline
				WatDiv $100$M	& $\mathbf{40.4}$	& $42.0$ 		& $71.4$	& $210.3$	& $96.4$	& $62.7$	& $767.2$	\\ \hline
		\end{tabular}
	}
	\caption{Query execution time, geometric mean (milliseconds)~\cite{AlucICDE2015}}
	\label{tab:evaluation:watdiv:overview}
\end{table}

\begin{figure}[t]
	\centering
	\begin{subfigure}[b]{0.235\textwidth}
		\includegraphics[width=\textwidth]{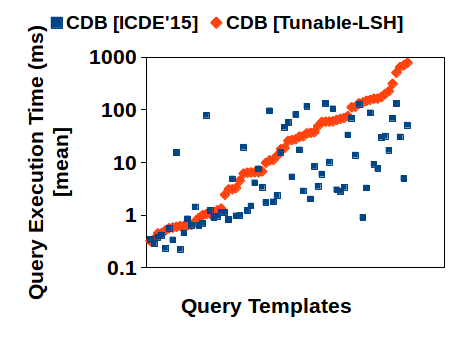}
		\caption{WatDiv $10$M triples}
		\label{fig:evaluation:cdb:detailed:small}
	\end{subfigure}
	\begin{subfigure}[b]{0.235\textwidth}
		\includegraphics[width=\textwidth]{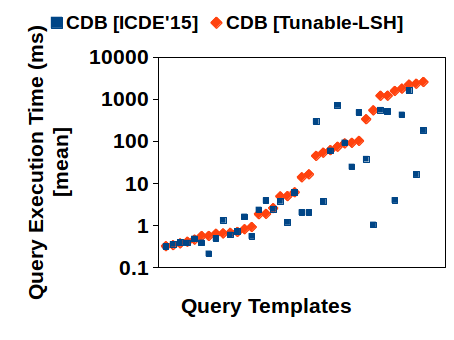}
		\caption{WatDiv $100$M triples}
		\label{fig:evaluation:cdb:detailed:large}
	\end{subfigure}
	\caption{Comparison of chameleon-db implemented using a hierarchical clustering algorithm and with \tunableLSH}
	\label{fig:evaluation:cdb:detailed}
\end{figure}

The first experiment evaluates \tunableLSH with\-in \emph{chameleon-db}, which is our prototype RDF data management system~\cite{AlucUW2013}.
In particular, in earlier work, we had introduced a hierarchical clustering algorithm for grouping RDF triples into what we call \emph{group-by-query clusters}~\cite{AlucICDE2015}.
In this evaluation, we replace that hierarchical clustering algorithm with \tunableLSH, and study its implications on the end-to-end query performance,
	keeping the same experimental configuration.
For completeness, we quote our description of the experimental setup from our previous paper:

``For our evaluations, we [primarily] use the Waterloo SPARQL Diversity Test Suite (WatDiv)
because it facilitates the generation of test cases that are far more diverse than any of the existing benchmarks~\cite{AlucISWC2014}.
In this regard, we use the WatDiv \emph{data generator} to create two datasets:
one with $10$ million RDF triples and another with $100$ million RDF triples
(we observe that systems under test (SUT) load data into main memory on the smaller dataset whereas at $100$M triples, SUTs perform disk I/O).
Then, using the WatDiv \emph{query template generator}, we create $125$ query templates 
and instantiate each query template with $100$ queries, thus, obtaining $12500$ queries.\footnote{\url{http://db.uwaterloo.ca/watdiv/stress-workloads.tar.gz}}"~\cite{AlucICDE2015}

We compare our approach with chameleon-db implemented with the hierarchical clustering algorithm (abbreviated CDB [ICDE'15]) and 
``five popular systems, namely, RDF-3x~\cite{NeumannVLDBJ2010}, MonetDB~\cite{IdreosIEEEDEB2012}, 4Store~\cite{HarrisSSWS2009} and 
Virtuoso Open Source (VOS) versions $6.1$~\cite{ErlingNKNM2009} and $7.1$~\cite{ErlingIEEEDEB2012}.
RDF-3x follows the single-table approach and creates multiple indexes;
MonetDB is a column-store, where RDF data are represented using vertical partitioning~\cite{AbadiVLDBJ2009};
and the last three systems are industrial systems.
Both 4Store and VOS group and index data primarily based on RDF predicates, but
VOS 6.1 is a row-store whereas VOS 7.1 is a column-store.
We configure these systems so that they make as much use of the available main memory as possible."~\cite{AlucICDE2015}

``We evaluate each system independently on each query template.
Specifically, for each query template, we first warm up the system by executing the workload for that query template once (i.e., $100$ queries).
Then, we execute the same workload five more times (i.e., $500$ queries).
We report average query execution time over the last five workloads."~\cite{AlucICDE2015}

``Our prototype starts with a completely segmented clustering, where each cluster consists of a single triple."~\cite{AlucICDE2015}
However, ``after the execution of the $100^{th}$ query, we allow the storage advisor to compute a better group-by-query clustering"~\cite{AlucICDE2015} 
using either the hierarchical clustering algorithm in~\cite{AlucICDE2015} or \tunableLSH.

Our experiments indicate that on average, the time to compute the group-by-query clusters has decreased by an order of magnitude with the introduction of \tunableLSH.
For example, for the $100$M triples dataset, it took $317.6$ milliseconds on (geometric) average to compute the group-by-query clusters using the hierarchical clustering algorithm in~\cite{AlucICDE2015},
	whereas with \tunableLSH, it takes only about $26.1$ milliseconds.
This is due to the approximate nature of \tunableLSH.
As shown in Table~\ref{tab:evaluation:watdiv:overview} and in Fig.~\ref{fig:evaluation:cdb:detailed}, this approximation has a slight impact on query performance, 
	but for the $100$M triples dataset, CDB is still significantly faster than the other RDF data management systems.
There is one apparent reason for this:
\tunableLSH is an approximate method, and therefore, the generated group-by-query clusters are not perfect.
To verify this hypothesis, we studied the logs generated during our experiments, which revealed the following:
using the group-by-query clustering in~\cite{AlucICDE2015}, 
	chameleon-db's query engine was able to execute $64.8\%$ of the queries without any decomposition (a property that chameleon-db's query optimizer is trying to achieve~\cite{AlucICDE2015}),
	whereas, the group-by-query clustering computed using \tunableLSH resulted in only $27.1\%$ of the queries to be executed without decomposition.
Of course, it is possible to improve chameleon-db's query optimizer, but that is a topic for future research.
%however, the point is that \tunableLSH will still generate approximate clusters.

This trade-off between the clustering overhead and the query execution time suggests that 
%	both the hierarchical clustering algorithm in~\cite{AlucICDE2015} and \tunableLSH have their own target use cases.
%More specifically, if the workloads can be predicted and sampled upfront, then physical clustering can take place once using the hierarchical algorithm in~\cite{AlucICDE2015}.
%On the other hand, 
for RDF workloads that are too dynamic to be predicted and sampled upfront, it might be desirable to have frequent clustering steps, in which case, using \tunableLSH is a much better option because of its lower overhead.

\subsection{Self-Clustering Hashtable}

\begin{figure*}
	\centering
	\begin{subfigure}[b]{0.31\textwidth}
		\includegraphics[width=\textwidth]{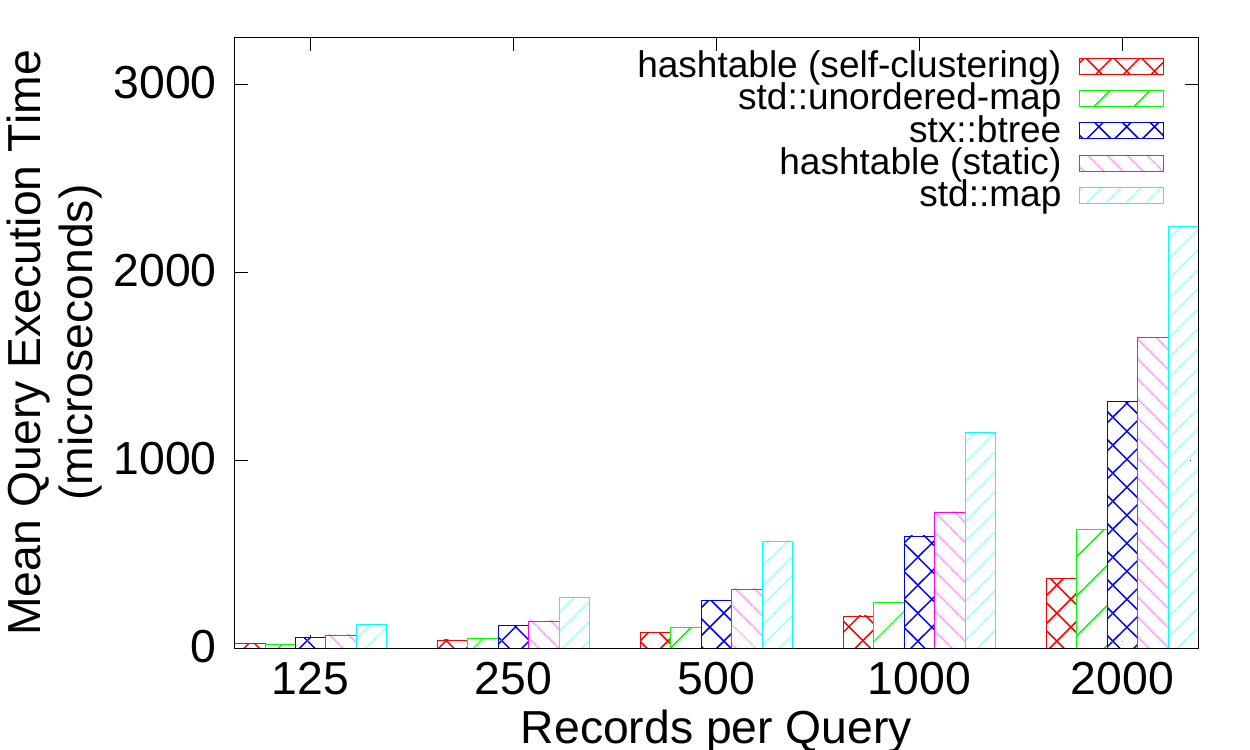}
		\caption{Random Access (All Data Structures) -- Control how loaded the workloads are}
		\label{fig:evaluation:hashtable:compare-all}
	\end{subfigure}
	\begin{subfigure}[b]{0.31\textwidth}
		\includegraphics[width=\textwidth]{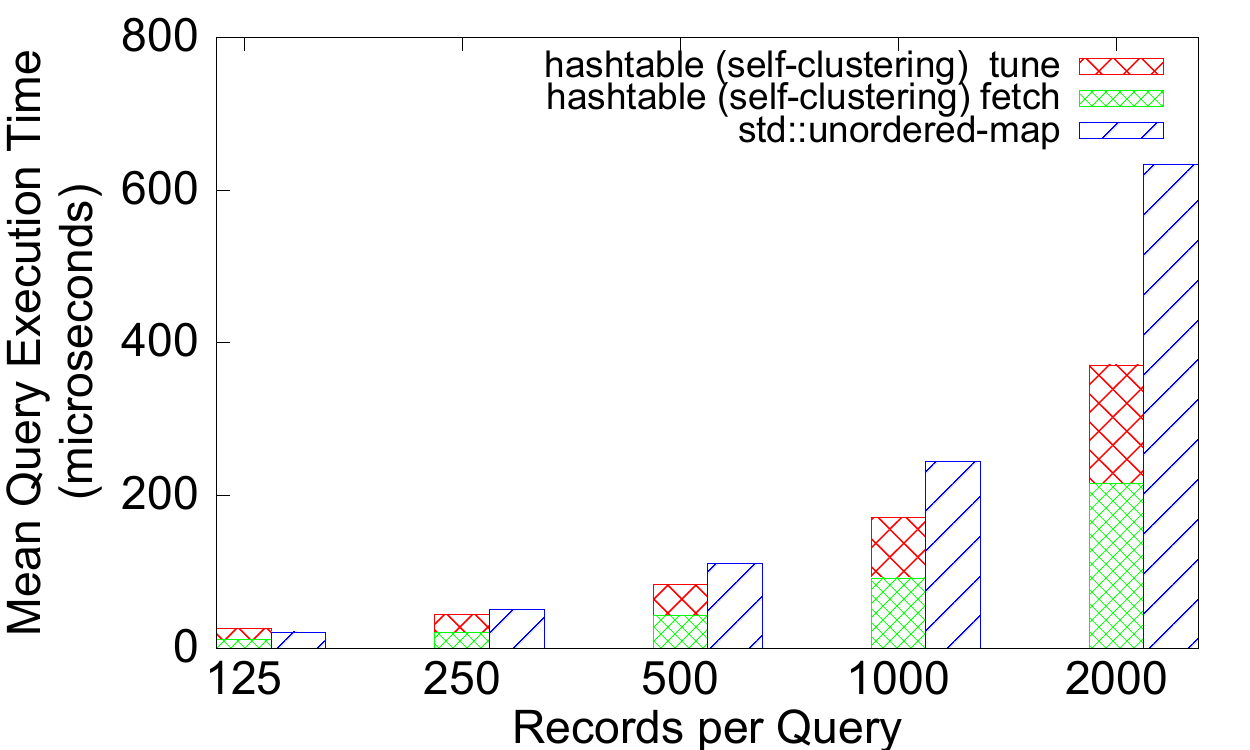}
		\caption{Random Access -- Control how loaded the workloads are}
		\label{fig:evaluation:hashtable:increasing-coverage}
	\end{subfigure}
	\begin{subfigure}[b]{0.31\textwidth}
		\includegraphics[width=\textwidth]{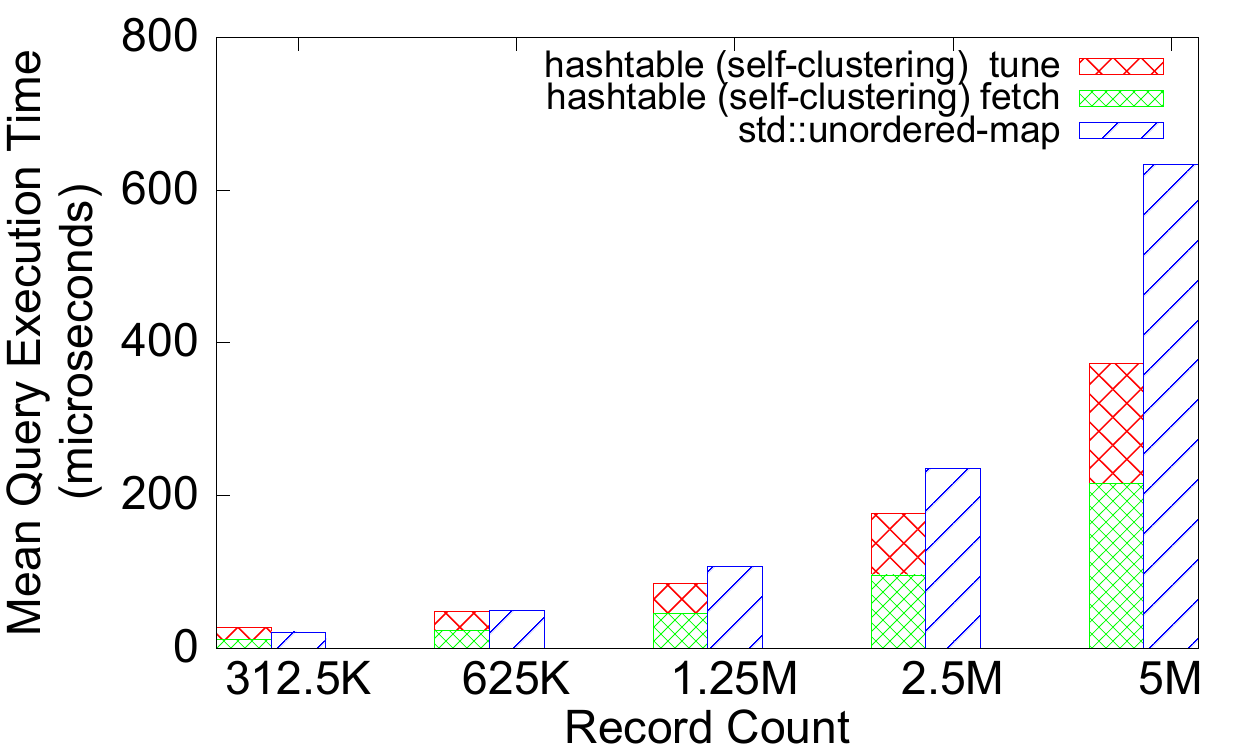}
		\caption{Random Access -- Control record count, keep record size constant at $128$ bytes}
		\label{fig:evaluation:hashtable:increasing-record-count}
	\end{subfigure}

	\begin{subfigure}[t]{0.31\textwidth}
		\includegraphics[width=\textwidth]{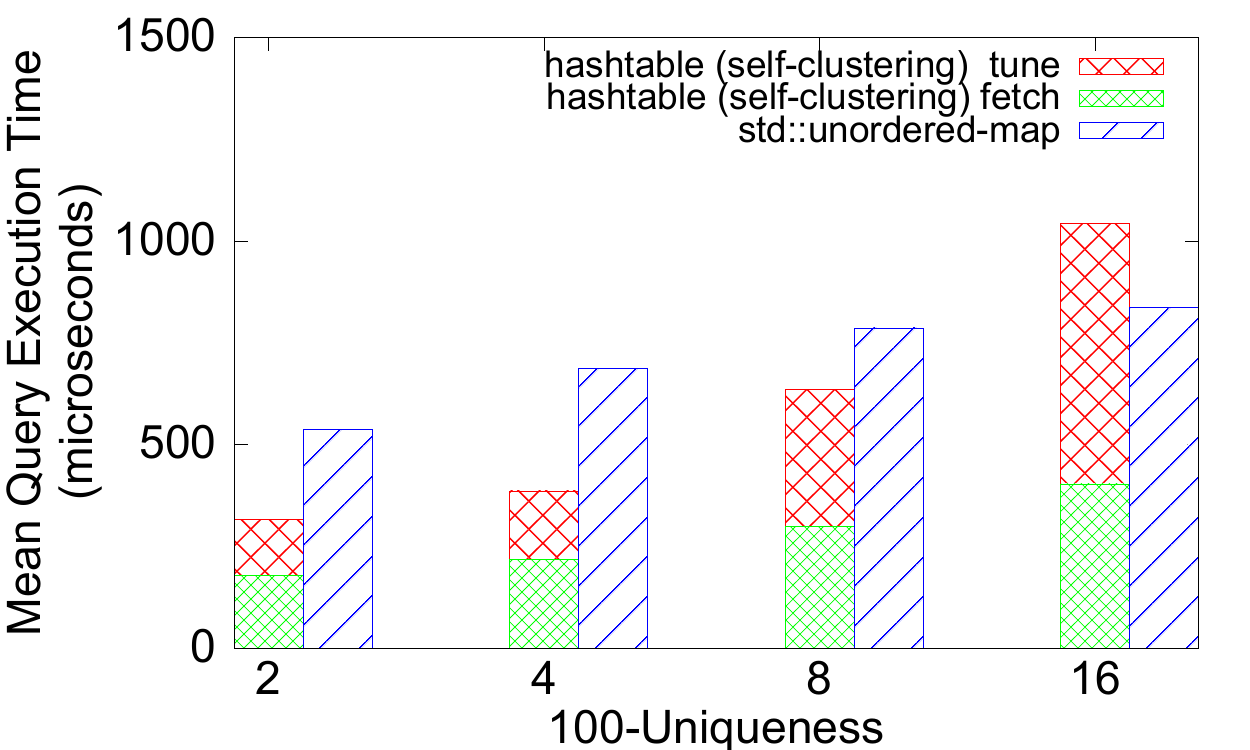}
		\caption{Random Access -- Control workload dynamism}
		\label{fig:evaluation:hashtable:increasing-uniqueness}
	\end{subfigure}
	\begin{subfigure}[t]{0.31\textwidth}
			\includegraphics[width=\textwidth]{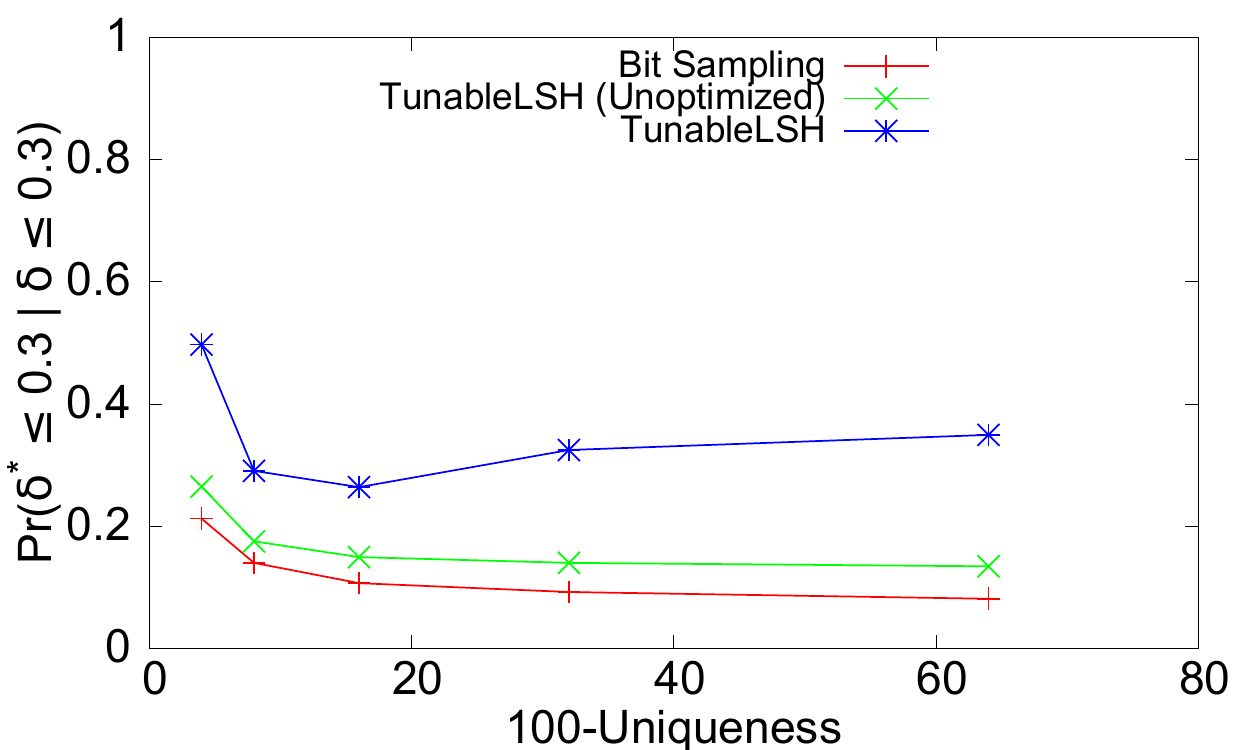}
			\caption{Sensitivity analysis of \tunableLSH -- Control workload dynamism}
			\label{fig:evaluation:tlsh:increasing-uniqueness}
	\end{subfigure}
	\begin{subfigure}[t]{0.31\textwidth}
			\includegraphics[width=\textwidth]{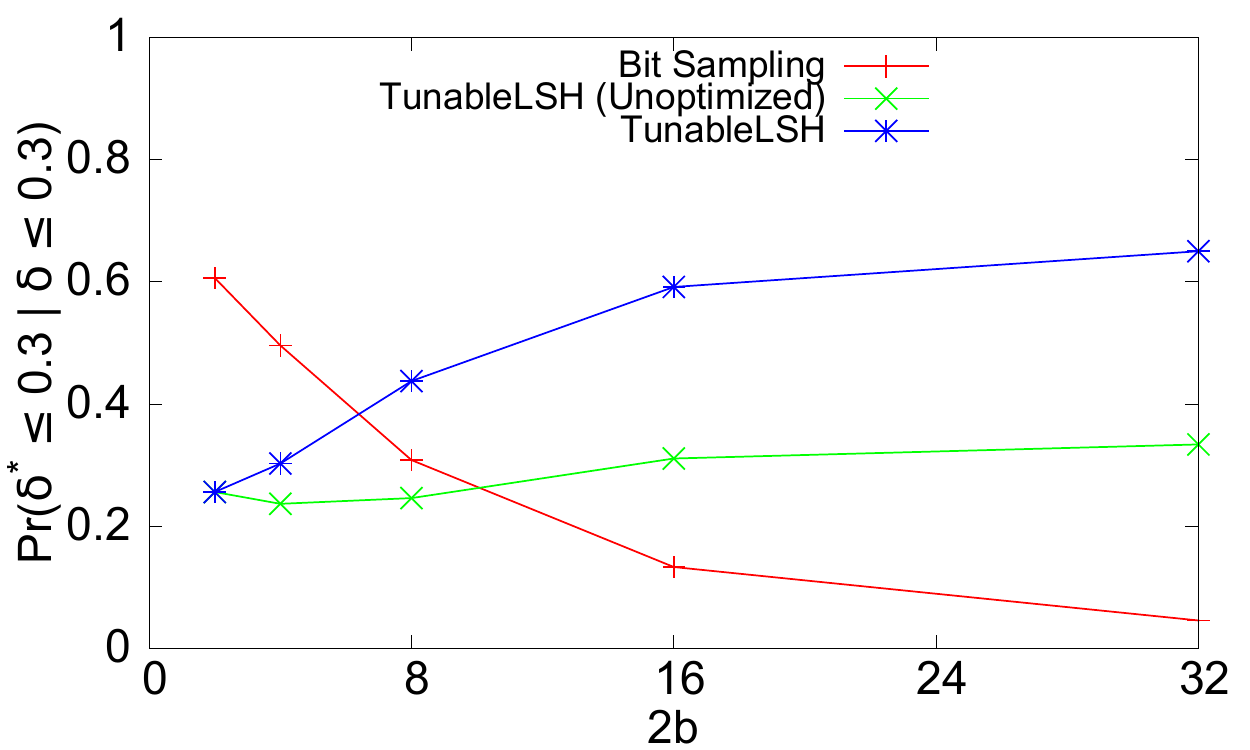}
			\caption{Sensitivity analysis of \tunableLSH -- Control \tunableLSH parameter $2b$}
			\label{fig:evaluation:tlsh:increasing-b}
	\end{subfigure}
	\caption{Experimental evaluation of \tunableLSH in a self-clustering hashtable and the sensitivity analysis of \tunableLSH}
	\label{fig:evaluation:hashtable}
\end{figure*}

The second experiment evaluates an in-memory hash\-table that we developed 
	that uses \tunableLSH to dynamically cluster re\-cords in the hashtable.
Hashtables are commonly used in RDF data management systems.
For example, the dictionary in an RDF data management system, 
	which maps integer identifiers to URIs or literals (and vice versa),
	is often implemented as a hashtable~\cite{WilkinsonHPL2006,AbadiVLDBJ2009,ErlingIEEEDEB2012}.
Secondary indexes can also be implemented as hashtables, 
	whereby the hashtable acts as a key-value store and 
	maps tuple identifiers to the content of the tuples.
In fact, in our own prototype RDF system, \emph{chameleon-db},
	all the indexes are secondary (dense) indexes
	because instread of relying on any sort order inherent in the data,
	we rely on the notion of group-by-query clusters, 
	in which RDF triples are ordered purely based on the workload~\cite{AlucUW2013,AlucICDE2015}.

The hashtable interface is very similar 
	to that of a standard hash\-table;	
	except that users are given the option to mark the beginning and end of queries.
This information is used to dynamically cluster records such that 
	those that are co-accessed across similar sets of queries also become physically co-located.
All of the clustering and re-clustering is transparent to the user, 
	hence, we call this the \emph{self-clustering hashtable}.

The self-clustering hashtable has the following advantages and disadvantages:
Compared to a standard hashtable that tries to avoid hash-collisions, 
	it deliberately co-locates records that are accessed together.
If the workloads favour a scenario in which many records are frequently accessed together, then 
	we can expect the self-clus\-tering hashtable to have improved fetch times 
	due to better CPU cache utilization, prefetching, etc.~\cite{AilamakiVLDB1999}. 
On the other hand, these optimizations come with three types of overhead.
First, every time a query is executed, \tunableLSH needs to be updated 
	(cf., Algorithms~\ref{alg:overview:tune} and~\ref{alg:online}).
Second, compared to a standard hashtable in which the physical address 
	of a record is determined solely using the underlying hash function (which is deterministic throughout the entire workload),
	in our case, the physical address of a record needs to be maintained dynamically
	because the underlying hash function is not deterministic (i.e., it is also changing dynamically throughout the workload).
Consequently, there is the overhead of going to a lookup table and retrieving the physical address of a record.
Third, physically moving records around in the storage system takes time---in fact, this is often an expensive operation.
Therefore, the objective of this set of experiments is twofold:
	\begin{inparaenum}[(i)]
		\item to evaluate the circumstances under which the self-clustering hashtable outperforms other popular data structures, and
		\item to understand when the tuning overhead may become a bottleneck.
	\end{inparaenum}
Consequently, we report the end-to-end query execution times, and if necessary,
	break it down into the time to 
	\begin{inparaenum}[(i)]
		\item \emph{fetch} the records, and 
		\item \emph{tune} the data structures 
			(which includes all types of overhead listed above).
	\end{inparaenum}

In our experiments, we compare the self-clustering hashtable to popular implementations of three data structures.
Specifically, we use:
	\begin{inparaenum}[(i)]
		\item \emph{std\texttt{::}unordered\_map}\footnote{\url{http://www.cplusplus.com/reference/unordered\_map/unordered\_map/}}, 
					which is the C\texttt{++} standard library implementation of a hashtable,
		\item \emph{std\texttt{::}map}\footnote{\url{http://www.cplusplus.com/reference/map/map/}}, 
					which is the C\texttt{++} standard library implementation of a red-black tree, and
		\item \emph{stx\texttt{::}\-btree}\footnote{\url{https://panthema.net/2007/stx-btree/}}, 
					which is an open source in-memory B\texttt{+} tree implementation.
	\end{inparaenum}
As a baseline, we also include a static version of our hashtable, i.e., one that does not rely on \tunableLSH.

We consider two types of workloads: 
	one in which records are accessed \emph{sequentially} and 
	the other in which records are accessed \emph{randomly}.
Each workload consists of $3000$ queries that are synthetically generated using WatDiv~\cite{AlucISWC2014}.
For each data structure, we measure the end-to-end workload execution time and compute the mean query execution time by dividing the total workload execution time by the number of queries in the workload.

Queries in these workloads consist of changing query access patterns, and in different experiments, we control different parameters such as the number of records that are accessed by queries on average, the rate at which the query access patterns change in the workload, etc.
We repeat each experiment $20$ times over workloads that are randomly generated with the same characteristics (e.g., average number of records accessed by each query, how fast the workload changes, etc.) and report averages across these $20$ runs.
We do not report standard errors because they are negligibly small and they do not add significant value to our results.

For the sequential case, \emph{stx\texttt{::}btree} and \emph{std\texttt{::}map} outperform the hashtables, 
	which is expected because once the first few records are fetched from main-memory,
	the remaining ones can already be prefetched into the CPU cache (due to the predictability of the sequential access pattern).
Therefore, for the remaining part, we focus on the random access scenario, 
	which is more common in RDF data management systems, 
	and which can be a bottleneck even in systems like RDF-3x~\cite{NeumannVLDBJ2010} 
	that have clustered indexes over all permutations of attributes.
For more examples and a thorough explanation, we refer the reader to~\cite{AlucPVLDB2014}.

In this experiment, we control the number of records that a query needs to access (on average), 
	where each record is $128$ bytes.
Fig.~\ref{fig:evaluation:hashtable:compare-all} compares all the data structures with respect to their end-to-end (mean) query execution times.
Three observations stand out:
First, in the random access case, the self-clustering hashtable as well as the standard hashtable perform much better than the other data structures, 
	which is what would be expected.
This observation holds also for the subsequent experiments, therefore, for presentation purposes, 
	we do not include these data structures in Fig.~\ref{fig:evaluation:hashtable:increasing-coverage}--\ref{fig:evaluation:hashtable:increasing-uniqueness}.
Second, the baseline static version of our hashtable (i.e., without \tunableLSH) performs 
	much worse than the standard hashtable, even worse than a B\texttt{+} tree.
This suggests that our implementation can be optimized further, 
	which might improve the performance of the self-clustering hashtable as well (this is left as future work).
Third, as the number of records that a query needs to access increases,
	the self-clustering hashtable outperforms all the other data structures,
	which verifies our initial hypothesis.

For the same experiment above, 
	Fig.~\ref{fig:evaluation:hashtable:increasing-coverage} focuses on the self-clus\-tering hashtable versus the standard hashtable, and
	illustrates why the performance improvement is higher (for the self-clustering hash\-table) 
	for workloads in which queries access more records.
Note that while the \emph{fetch} time of the self-clustering hashtable scales proportionally 
	with respect to std\texttt{::}unordered\_map, 
	the \emph{tune} overhead is proportionally much lower for workloads 
	in which queries access more records.
This is because with increasing ``records per query count", 
	records can be re-located in batches across the pages in main-memory
	as opposed to moving individual records around.

Next, we keep the average number of records that a query needs to access constant at $2000$, 
	but control the number of records in the database.
As in the previous experiment, each record is $128$ bytes.
As illustrated in Fig.~\ref{fig:evaluation:hashtable:increasing-record-count}, increasing the number of records in the database
	(i.e., scaling-up) favours the self-clustering hashtable.
The reason is that, when there are only a few records in the database, 
	the records are likely clustered to begin with.
We repeat the same experiment, but this time, by controlling the record size and keeping the database size constant at $640$ megabytes.
Surprisingly, the relative improvement with respect to the standard hashtable remains more or less constant,
	which indicates that the improvement is largely dominated by the size of the database, 
	and increasing it is to the advantage of the self-clustering hashtable.

Finally, we evaluate how sensitive the self-clustering hashtable is to the dynamism in the workloads.
Note that for the self-clustering hashtable to be useful at all, the workloads need to be predictable---at least to a certain extent.
That is, if records are physically clustered but are never accessed in the future, then all those clustering efforts are wasted.
To verify this hypothesis, we control the expected number of query clusters 
	(i.e., queries with similar but not exactly the same access vectors) 
	in any $100$ consecutive queries in the workloads that we generate.
Let us call this property of the workload, its \emph{$100$-Uniqueness}.
Fig.~\ref{fig:evaluation:hashtable:increasing-uniqueness} illustrates how the \emph{tuning} overhead starts to become a bottleneck 
	as the workloads become more and more dynamic, to the extent of being completely unique, i.e., each query accesses a distinct set of records.

%We are planning to make the code of the self-clustering hashtable 
%	as well as the workload generator publicly available;
%	the latter, as part of WatDiv~\cite{AlucISWC2014}.

\subsection{Sensitivity Analysis of Tunable-LSH}

In the final set of experiments, we evaluate the sensitivity of \tunableLSH in isolation, that is,
	without worrying about how it affects physical clustering, and compare it to three other hash functions:
	\begin{inparaenum}[(i)]
		\item a standard non-locality sensitive hash function\footnote{\url{http://en.cppreference.com/w/cpp/utility/hash}},
		\item bit-sampling, which is known to be locality-sensitive for Hamming distances~\cite{IndykSTOC98}, and
		\item \tunableLSH without the optimizations discussed in Section~\ref{sec:details}.
	\end{inparaenum}
These comparisons are made across workloads with different characteristics (i.e., dense vs.~sparse, dynamic vs.~stable, etc.) 
	where parameters such as 
	the average number of records accessed per query and
	the expected number of query clusters within any $100$-consecutive sequence of queries in the workload are controlled.

Our evaluations indicate that \tunableLSH generally outperforms its alternatives.
Due to space considerations, we cannot present all of our results in detail.
Therefore, we will summarize our most important observations.
%We are planning to make our C\texttt{++} implementation of \tunableLSH, as well as the scripts for running these experiments publicly available.

Fig.~\ref{fig:evaluation:tlsh:increasing-uniqueness} shows how 
	the probability that \emph{the evaluated hash functions place records with similar utilization vectors to nearby hash values}
	changes as the workloads become more and more dynamic.
In computing these probabilities, both the original distances (i.e., $\delta$) and the distances over the hashed values (i.e., $\delta^{*}$) 
	are normalized with respect to the maximum distance in each geometry.
As illustrated in Fig.~\ref{fig:evaluation:tlsh:increasing-uniqueness}, \tunableLSH achieves higher probability even when the workloads are dynamic.
The unoptimized version of \tunableLSH behaves more or less like a static locality-sensitive hash function, such as bit sampling, which is an expected result because
	\tunableLSH cannot achieve high accuracy
	without the workload-sensitive arrangement introduced in Section~\ref{sec:details}.
It is also important to emphasize that even in that case \tunableLSH is no worse than a standard LSH scheme, which is aligned with the theorems in Section~\ref{sec:details:properties}.
We have not included the results on the standard non-locality sensitive hash function,
	because, as one might guess, it has a probability distribution that is completely unparalleled to our clustering objectives.

Fig.~\ref{fig:evaluation:tlsh:increasing-b} demonstrates how the choice of $b$ (or $2b$ as described in Section~\ref{sec:details:reset}) affects the accuracy of \tunableLSH.
Having a higher $b$ implies less and less undesirable collisions of query access vectors, hence, a higher accuracy.
On the other hand, for bit sampling, the ideal number of samples is equal to the query clusters in the workload, thus, increasing $b$, 
	which corresponds to the number of bits that are sampled, 
	might result in oversampling and therefore, lower accuracy.
For example, consider two record utilization vectors $1001$ and $0001$ with Hamming distance $1$.
If only $1$ bit is sampled, there is $\frac{3}{4}$ probability that these two vectors will be hashed to the same value.
On the other hand, if $2$ bits are sampled, the probability drops to $\frac{1}{2}$.

\section{Conclusions and Future Work}
\label{sec:conclusions}

In this paper, we introduce \tunableLSH, which is a locality-sensitive hashing scheme,
and demonstrate its use in clustering records in an RDF data management system.
In particular, we keep track of the fragmented records in the database and use \tunableLSH to decide, 
	in constant-time, where a record needs to be placed in the storage system.
\tunableLSH takes into account the most recent query access patterns over the database, 
	and uses this information to auto-tune such that records that are accessed across similar sets of queries
	are hashed as much as possible to the same or nearby pages in the storage system.
This property distinguishes \tunableLSH from existing locality-sensitive hash functions, which are static.
Our experiments with 
(i) a version of our prototype RDF data management system, \emph{chame\-leon-db}, that uses \tunableLSH,
(ii) a hashtable that relies on \tunableLSH to dynamically cluster its records, and
(iii) workloads that rigorously test the sensitivity of \tunableLSH verify the potential benefits of \tunableLSH.

As future work, it would be beneficial to answer the following questions.
First, the assumption that the last $k$ queries are representative of the future queries in the workload can be relaxed.
As outlined in~\cite{AlucPVLDB2014}, the issue of deciding ``when and based on what information to tune the physical design" of our system still remains an open problem.
Second, as our experiments indicate, query optimization in \emph{chame\-leon-db} has significant room for improvement.
We need techniques that can handle more approximate group-by-query clusters such as those generated by \tunableLSH.
Third, we believe that \tunableLSH can be used in a more general setting than just RDF systems.
In fact, it should be possible to extend the idea of the self-clustering in-memory hashtable that we have implemented to a more general, distributed key-value store.

{
%\scriptsize
\bibliographystyle{abbrv}
\bibliography{publications.bib,hashing-and-caching-paper.bib}
}

%%%\appendix
%%%\input{sections/appendix}

\end{document}